%% file: su2ha_spec_2.tex
\documentstyle[12pt,epsf]{article}
\newcommand{\beq}{\begin{equation}}
\newcommand{\eeq}{\end{equation}}
\newcommand{\bea}{\begin{eqnarray}}
\newcommand{\eea}{\end{eqnarray}}
\newcommand{\bi}{\begin{itemize}}
\newcommand{\ei}{\end{itemize}}

\newcommand{\tr}{{\rm Tr\,}}
\newcommand{\nn}{\nonumber}
\newcommand{\fr}[2]{{\frac{#1}{#2}}}

\newcommand{\lambdamsbar}{\Lambda_{\overline{\rm MS}}}

\renewcommand{\vec}[1]{{\bf #1}}
\def\mb#1         {\mbox{\boldmath $#1$}}

\def\lsi{\raise0.3ex\hbox{$<$\kern-0.75em\raise-1.1ex\hbox{$\sim$}}}
\def\gsi{\raise0.3ex\hbox{$>$\kern-0.75em\raise-1.1ex\hbox{$\sim$}}}

\newcommand{\mcempty}{\multicolumn{2}{c}{\mbox{}}}

\newcommand{\mcemptyr}{\multicolumn{2}{c|}{\mbox{}}}
\newcommand{\mcemptyll}{\multicolumn{2}{||c}{\mbox{}}}
\newcommand{\mcemptyrr}{\multicolumn{2}{c||}{\mbox{}}}

\begin{document}

\begin{titlepage}
\begin{flushright}
Edinburgh 99/8\\
CERN-TH/99-253\\
hep-lat/9908041\\
\hspace{1em}\\
August 1999
\end{flushright} \begin{centering} \vfill

{\bf THE SPECTRUM OF THE THREE-DIMENSIONAL \\
ADJOINT HIGGS MODEL AND HOT SU(2) GAUGE THEORY}
\vspace{0.8cm}

A. Hart$^{\rm a}$ and 
O. Philipsen$^{\rm b}$

\vspace{0.3cm}
{\em $^{\rm a}$%
Department of Physics and Astronomy, University of Edinburgh,\\
Edinburgh EH9 3JZ, Scotland, UK\\}
\vspace{0.3cm}
{\em $^{\rm b}$%
Theory Division, CERN, CH-1211 Geneva 23,
Switzerland\\}

\vspace{0.7cm}
{\bf Abstract.}
\end{centering}
We compute the mass spectrum of the SU(2) adjoint Higgs model in
$2+1$ dimensions at several points located in the (metastable)
confinement region of its phase diagram.  We find a dense spectrum
consisting of an almost unaltered repetition of the glueball spectrum
of the pure gauge theory, and additional bound states of adjoint
scalars.  For the parameters chosen, the model represents the
effective finite temperature theory for pure SU(2) gauge theory in
four dimensions, obtained after perturbative dimensional reduction.
Comparing with the spectrum of screening masses obtained in recent
simulations of four-dimensional pure gauge theory at finite
temperature, for the low lying states we find quantitative agreement
between the full and the effective theory for temperatures as low as
$T=2T_c$. This establishes the model under study as the correct
effective theory, and dimensional reduction as a viable tool for the
description of thermodynamic properties.  We furthermore compare the
perturbative contribution $\sim gT$ with the non-perturbative
contributions $\sim g^2T$ and $\sim g^3T$ to the Debye mass. The
latter turns out to be dominated by the scale $g^2T$, whereas higher
order contributions are small corrections.

\vspace{0.3cm}\noindent

\vfill


\vfill

\noindent
CERN-TH/99-253\\
August 1999

\vfill

\end{titlepage}

\section{Introduction}

The thermodynamic properties of non-Abelian field theories at finite
temperature are known to be non-perturbative due to the confining
nature of the Matsubara zero mode sector.  Lattice simulations of the
full problem are numerically expensive and especially difficult if
fermions are present. A way to circumvent both problems is to
perturbatively integrate out the non-zero Matsubara and momentum modes
of order $\sim T$ and higher to arrive at a purely bosonic,
three-dimensional effective theory describing the static quantities of
the original theory \cite{dr}.  Three-dimensional gauge theories are
non-perturbative, but readily accessible by lattice simulations. This
programme has been successfully carried out in studies of the
electroweak phase transition, where the phase structure and nature
of the transition were computed from the dimensionally reduced theory,
both by numerical \cite{ewn} and analytical \cite{ewa} methods.
The results have been found to agree to remarkable
accuracy with the answers obtained recently from four-dimensional simulations
\cite{4dew}.

In view of the experimental search for the quark gluon plasma in heavy
ion collisions at RHIC and LHC it is of particular interest to see to
what extent this programme can be carried over to QCD. The reduction
step for a SU(N) gauge theory with $N_f$ flavours of fermions has been
carried out to two loops in \cite{ad} (see also \cite{bra}).  The
effective theory obtained in this case is a three-dimensional SU(N)
adjoint Higgs theory, with the scalars corresponding to the electric
gauge potential, $A_0$, in the unreduced theory. In the framework of
dimensional reduction $N_f$ only modifies the parameters of the
effective, bosonic theory.  To be specific, and to compare with
existing four-dimensional work, we discuss in this study the simplest
case of $N_f=0, N=2$.

Simulations of the three-dimensional SU(2) adjoint Higgs model 
revealed that this theory has a phase diagram with continuously connected
Higgs and confinement regions 
\cite{usad,ad}, 
which are partially separated by a line of first order phase
transitions.  This is in contrast to the four-dimensional SU(2)
Yang-Mills theory at finite T, which displays a second order phase
transition \cite{4dpt}. Hence the deconfinement phase transition
cannot be described by the effective theory. 
Based on the apparent convergence properties of the reduction step, however,
it has been argued \cite{ad} that the effective theory still 
yields the correct
description of static correlation functions at temperatures
sufficiently above the critical temperature, $T_c$, and hence valuable
information about the plasma state.

In recent publications \cite{dg,dg2} it was established that
dimensional reduction indeed works in the sense that the spectrum
measured in four-dimensional simulations at finite temperature can be
classified by the symmetry group of a purely three-dimensional theory.
This does not yet determine the precise form of the effective theory.
Earlier simulations of the SU(2) adjoint Higgs model have concluded it
to be a good effective theory by comparing the static potential \cite{rei1}
and gauge-fixed propagators \cite{kar1} between the full and the effective
theory. The question of the dynamical r{\^o}le of the $A_0$ and whether
there is a further hierarchy $gT\ll g^2T$ in the reduced 
theory, however, is still not precisely answered.
The purpose of the present paper is to quantitatively test the
agreement of a wide range of gauge-invariant 
correlation functions between the full and the
effective theory at temperatures not much larger than the critical
temperature.  We compute correlation functions of the
three-dimensional SU(2) adjoint Higgs model on the
lattice and extract the corresponding screening masses for various
quantum number channels. These are then compared with the
four-dimensional calculation of finite temperature SU(2) Yang-Mills
theory in \cite{dg2}. 
We find indeed that dimensional
reduction works remarkably well, even at the quantitative level.
Furthermore, we are able to identify the lightest gauge-invariant state
to consist of $A_0$, which hence may not be integrated out.

The continuum action of the SU(2) adjoint Higgs model is given by
\begin{equation} \label{actc}
        S = \int d^{3}x \left\{ \frac{1}{2} \tr(F_{ij}F_{ij})
        +\tr(D_{i}\varphi D_{i} \varphi) +m_3^{2} \tr(\varphi \varphi)
        +\lambda_3(\tr(\varphi \varphi))^{2} \right\} ,
\end{equation}
where $F_{ij}=\partial_{i}A_{j}-\partial_{j}A_{i}
 +ig[A_i,A_j]$,
$D_i\varphi = \partial_i\varphi + ig_3[A_i,\varphi]$
and $F_{ij},A_i$, and $\varphi$ are all traceless $2\times 2$
Hermitian matrices ($\varphi=\varphi^{a}\sigma_{a}/2$, {\it etc.}).
The physical properties of the theory are fixed by  
the two dimensionless ratios
\beq
x=\frac{\lambda_3}{g_3^2}, \quad y=\frac{m_3^2}{g_3^4}.
\eeq
In the framework of dimensional reduction these parameters are
completely determined by the four-dimensional gauge coupling, $g^2$,
and the temperature, $T$. Since $g^2$ is a running coupling its value
is in turn fixed by a renormalisation scale $\lambdamsbar$.  The
details of the derivation of $x,y$ to two loops can be found in
\cite{ad}.
Here we merely quote the results to furnish the connection between the
three-dimensional parameters and the four-dimensional, finite $T$
situation. Choosing the renormalisation scale as in \cite{ad} and
expressing $\lambdamsbar$ through the critical temperature as measured
on the lattice \cite{4dpt}, $T_c=1.23(11) \lambdamsbar$, one has
\beq \label{3d}
g_3^2=\frac{10.7668}{\ln(8.3T/T_c)} T, 
\quad x=\frac{0.3636}{\ln(6.6T/T_c)}, 
\quad y(x)=\frac{2}{9\pi^2 x}+\frac{1}{4\pi^2} + {\cal O}(x).
\eeq
With these equations, specifying $T/T_c$ completely 
fixes the parameters $x,y$ of the reduced model.

In this paper we study specifically the temperatures $T=2T_c$,
$T=4T_c$ and $T=5T_c$ (with the corresponding values of $x$ and $y$
given in Table~\ref{tab_params}). Note, however, that these parameters
lie in the Higgs phase of the model, whereas the connection to four
dimensional physics is only valid in the confinement phase \cite{rei,ad}.
The dimensional reduction programme would be futile were it not that
the confinement phase is metastable in this region of the phase
diagram.  Starting with a configuration in the confinement phase, we
may re-thermalise and make many measurements at the appropriate
parameters before a tunnelling transition occurs to the Higgs phase.
In practice, the phases are so strongly separated that no tunnelling
occurred in our simulations.

In order to discuss correlation functions of the effective theory, it
is most convenient to ignore the finite T context and to discuss the
properties of the $2+1$ dimensional Higgs model by itself, as we shall
do in the following two sections. This will also enable us to compare
the dynamics of the SU(2) adjoint Higgs model in its confining phase
with that of the $2+1$ dimensional pure gauge theory and fundamental
Higgs model, whose spectra are known in some detail \cite{mt,ptw}, and
thus to learn about some general features of three-dimensional
confinement.  In Sec.~2 we introduce the corresponding lattice action,
the operators considered and some technical details about our
numerical simulation. In Sec.~3 we present our detailed numerical
results for the spectrum of the model, including checks for finite
size effects and a numerical continuum extrapolation. Then we return
in Sec.~4 to an interpretation and discussion of our results in the
context of finite temperature pure gauge theory, before giving our
conclusions.

\section{Lattice methods}

\subsection{Action and parameters}

The discretised form of the action (\ref{actc}), with leading order
lattice spacing corrections of ${\cal O}(a)$, is
        \begin{eqnarray}
        S &=& \beta \sum_{x,i > j}\left(1-\frac{1}{2}
        \tr U_{ij}(x)\right)
        +2\sum_{x} \tr(\varphi(x)\varphi(x)) \nonumber \\
        &-&2\kappa \sum_{x,\mu}\tr(\varphi(x)U_{\mu}
        (x)\varphi(x+\hat{\mu}a)U^{\dagger}(x))
        +\lambda \sum_{x}(2 \tr(\varphi(x)\varphi(x))-1)^{2}.
        \label{lattice_action}
        \end{eqnarray}
(The lattice scalar field has been rescaled relative to the continuum.)
The parameters of the continuum and lattice theory are up to two loops
related by a set of equations specifying the lines of constant
physics in the lattice parameter space \{$\beta,\kappa,\lambda$\} \cite{lai95},
\bea
\beta&=&\frac{4}{ag_3^2}, \quad
\lambda=\frac{x\kappa^2}{\beta}, \nn\\
y&=& \frac{\beta^2}{8}
                \left(\frac{1}{\kappa}-3
                -\frac{2x\kappa}{\beta}\right) 
                +\frac{\Sigma\beta}{4\pi}
                     \left(1+\frac{5}{4}x\right) \nonumber\\
                &+&\frac{1}{16\pi^{2}}
                  \left[(20x-10x^2)
                  \left(\ln\left(\frac{3\beta}{2}\right)+0.09\right)
                  +8.7+11.6x\right].
\eea
where $\Sigma = 3.17591$.  For a given pair of continuum parameters
$x,y$ these equations determine the lattice parameters
$\kappa,\lambda$ as a function of lattice spacing, and hence govern
the approach to the continuum limit, $\beta\rightarrow \infty$.  It is
along such trajectories that we should observe a scaling of
dimensionful quantities such as the masses. Due to
superrenormalisability of the theory, the above perturbative relation
is exact in the continuum limit. Based on previous experience
\cite{usad} we expect it to be accurate for all $\beta > 6$.  As a
consequence of the presence of scalar fields in the action, the
leading order lattice spacing corrections to mass ratios should be
linear in $a$.

\subsection{Blocking of fields}

Since we work in the confinement region of the theory, the physical
states are expected to be bound states, which are spatially extended.
In order to improve the projection of any operator onto such states,
it is necessary to smear the operators
in the spatial plane so that their extension becomes
comparable to the size of the states. This can be achieved by using
``blocked'' field variables instead of the elementary ones in the
expressions for the operators. We use blocked link variables at 
blocking level $n$
as proposed in \cite{block},
\bea \label{lbl}
U_{i}^{(n)}(x)&=&\fr13
\left\{U_i^{(n-1)}(x)U_{i}^{(n-1)}(x+\hat{i}) \right.\nn\\
&&+\left.\sum_{j=\pm 1,j\neq i}^{\pm 2}
U^{(n-1)}_j(x)U_i^{(n-1)}(x+\hat{j})
U_i^{(n-1)}(x+\hat{i}+\hat{j})U_j^{(n-1)\dagger}(x+2\hat{i})\right \}.
\eea
For the scalars, we adapt the procedure employed in \cite{ptw}
to the case of adjoint fields,
\bea
\varphi^{(n)}(x)&=&\frac{1}{5}\left\{\varphi^{(n-1)}(x)
+\sum_{j=1}^{2}\left[
U_j^{(n-1)}(x)\varphi^{(n-1)}(x+\hat{\j})U_j^{(n-1)\dag}(x)\right.\right.\nn\\
&&\left.\left.\hspace*{3cm}+U_j^{(n-1)\dag}(x-\hat{\j})
\varphi^{(n-1)}(x-\hat{\j})U_j^{(n-1)}(x-\hat{\j})\right]\right\}\;.
\eea
Note that we will take correlations in the $\hat{3}$ direction and so
our blocking always remains within the $(1,2)$-plane to avoid spoiling
the transfer matrix of the theory and the positivity properties of the
correlation functions. In addition, we consider blocked scalar fields
$\varphi^{(n,j)}(x)$ with non-local contributions from one spatial
direction~$j$ only:
\bea
\varphi^{(n,j)}(x)=\frac{1}{3}\Big\{\varphi^{(n-1,j)}(x) & + &
U_j^{(n-1)}(x)\varphi^{(n-1,j)}(x+\hat{\j})U_j^{(n-1)\dag}(x))
\\
& + &U_j^{(n-1)\dag}(x-\hat{\j})\varphi^{(n-1,j)}(x-\hat{\j})
U_j^{(n-1)}(x-\hat{\j})\Big\},\;j=1,2.\nn
\label{asymblock}
\eea

\subsection{Basic operators}

The operators we use in our simulations are constructed from several basic
gauge-invariant operator types, which we now list
before specifying the quantum numbers.
We have operators involving only scalar fields or products of scalar
and gauge field variables,
\bea
R(x)&=&\tr(\varphi^2(x)) \nn\\ 
L_i(x)&=&\tr \left( \varphi(x)U_i(x)\varphi(x+\hat{\i})U_i^{\dag}(x) \right)\nn\\ 
A^+_{ij}(x)&=&\tr\left( U_i(x)
U_i(x+\hat{\i})\varphi(x+2\hat{\i})U^{\dag}_i(x+\hat{\i})U^{\dag}_i(x)
U_j(x)\varphi(x+\hat{\j})U^{\dag}_j(x)\right)\\
A^-_{ij}(x)&=&\tr\left( U_i(x)
U_i(x+\hat{\i})\varphi(x+2\hat{\i})U^{\dag}_i(x+\hat{\i})U^{\dag}_i(x)
U_j(x)\varphi(x+\hat{\j})U^{\dag}_j(x)
\varphi(x)\right)\,.\nn 
\eea
Other mixed operators are 
\bea \label{bop}
B_{||}(x)&=&\tr(\varphi(x) U_{ij}(x)),\nn\\
B_{\perp}(x)&=&\left[\tr T^a\left(U_{ij}(x)
-\varphi(x)\frac{\tr(\varphi(x) U_{ij}(x))}{\sqrt{\tr(\varphi^2(x))}}\right)\right]^2\,.
\eea
The naming of the operators becomes apparent when considering the continuum 
limit, in which $B_{||}(x)\sim \tr(\varphi(x) F_{ij}(x))$ gives 
the projection of the field strength along the scalar field and $B_{\perp}(x)$
is the projection perpendicular to it.

Further, we have loop operators constructed from link variables only,
\beq
 C_{ij}^{1\times1}(x)=\tr\left(
 U_i(x)U_j(x+\hat{\i})U^\dag_i(x+\hat{\j})U^\dag_j(x) \right),
 \quad i,j=1,2,\,i\not=j,
\eeq
and in addition to the elementary plaquette $C^{1\times1}$, we also
consider squares of size $2\times2$ as well as rectangles of size
$1\times2$, $1\times3$, $2\times3$. Hence, we have five versions of
operators consisting of closed loops of gauge fields, {\it viz.}
\beq
  C_{ij}^{1\times1},\;C_{ij}^{2\times2},\;C_{ij}^{1\times2},\;
  C_{ij}^{1\times3},\;C_{ij}^{2\times3}.
\eeq
Another pure gauge 
operator useful to probe the confining properties of the theory
is the Polyakov loop along a spatial direction $j$,
\beq
  P_j^{(L)}(x)=\tr\prod_{m=0}^{L-1}\,U_j(x+m\hat{\j}), \quad j=1,2.
\eeq
Apart from its relevance for confinement, the Polyakov loop also
plays an important r\^ole in the study of finite size effects.
Although single torelons
arising from correlations of the Polyakov loop $P_j^{(L)}$ can neither
contribute directly to correlations of operators $C^{1\times1},\ldots,
C^{2\times3}$, nor to our mixed operators, 
they may well couple indirectly on the grounds that they
share the quantum numbers with the infinite volume states.
Furthermore,
torelon-antitorelon pairs are known to give rise to 
finite-volume effects in glueball calculations\,\cite{cm87a}. Apart
from studying correlations of Polyakov loops {\it per se\/}, we have
thus constructed single-torelon and torelon-pair operators in the
$0^+_+$ and $2^+_+$ channels and studied their correlations as a
safeguard against finite-size effects.

\subsection{Quantum numbers}

Operators with definite quantum number assignments may be constructed
from the above operator types by taking linear combinations with
appropriate transformation properties, {\it i.e.} by applying
projection operators for the various irreducible represent\-ations of
the symmetry group of the lattice slice upon which the operators are
constructed.

In a three-dimensional Euclidian gauge theory, the slice is a plane with an
$SO(2)$ rotational symmetry and discrete $Z_2$ parity reflection and 
charge conjugation symmetries. 
Rotations are generated by $R(\theta)$ under which
operators of angular momentum $j$ gain a phase $\exp[{\rm i}j\theta]$.
Parity in two spatial dimensions reflects operators in one
coordinate axis;
$P:(x,y) \to (x,-y)$;
reflections in other axes are related to this by rotations.  Charge
conjugation acts as complex conjugation on the matrix valued fields,
$C:U_i(x) \to U^*_i(x), \varphi(x) \to \varphi^*(x)$.  Physical
states and operators are characterised by their eigenvalues under
these operations, the quantum numbers $J^{PC}$.  In contrast to the
fundamental Higgs model, there are no $C=-1$ states in the SU(2) pure
gauge theory or the SU(2) adjoint Higgs model. In this model SU(2)
charge conjugation is equivalent to a global gauge transformation, and
hence any gauge-invariant operator is even under this operation.  We
therefore omit the index $C$ from now on, implying that all states
have $C=+1$.

With an adjoint scalar field present, there is an additional $Z_2$
symmetry corresponding to reflections $\varphi \to -\varphi$, whose
eigenvalue we call $R$.  Clearly operators containing an even number
of scalar fields, such as $A^+_{ij}$, will couple to $R = +1$ states,
and those with odd, such as $A^-_{ij}$, to $R = -1$. We thus classify
our operators and the states they couple to by $J^P_R$.

On the lattice the symmetry group is broken from $SO(2) \otimes Z_2(P)
\otimes Z_2(R)$ to the point group $C^4_v \otimes Z_2(R)$, which
restricts the allowed rotations to $R(\theta_n)$ where $\theta_n = n
\pi / 2$. Strictly speaking, we should thus classify our states
by the irreducible representations of $C^4_v$ rather than by $J^P$.
Since we are really interested in continuum physics and, as we shall see,
our data reproduce the continuum symmetries within the statistical errors,
we prefer to keep the continuum notation.

The fact that the two-dimensional parity operator reflects only one spatial
direction, and hence the angular momentum operator changes sign under
$P$, has an important consequence for the spectrum: in $2+1$
dimensions all states with $J \neq 0$ come in degenerate pairs of
opposite parity (see, {\it e.g.}, \cite{mt}). This statement of
``parity doubling'' is based on the continuum rotation group, whereas
a square lattice with a finite volume only admits rotations in units
of $\pi/2$. Thus, although we should continue to observe parity
doubling of $1^+_R$ and $1^-_R$ on the lattice, the doubling in the
$J=2$ sector is lost, and will only be recovered in the infinite
volume limit. Comparing the masses of the $2^+_R$ and $2^-_R$ states
is thus a useful measure of whether our lattice is free from
discretisation and finite volume effects.  In this study all
simulations were done on lattices large enough that parity doubling
was, within statistical errors, in accordance with continuum, infinite
volume expectations.

An interesting feature of the SU(2) fundamental Higgs model is the
near replication of the pure gauge theory glueballs within its
spectrum.  Whereas one might expect the excitations of the former to
be a mixture of gluonic and scalar degrees of freedom, it is found
instead that within the spectrum there are some essentially purely
gluonic states.  Given their remarkable decoupling from the scalar
fields, it is then unsurprising that the masses of these match that of
the pure gauge model almost exactly \cite{ptw}.  One of our main
interests is to study whether the glueball spectrum survives the
addition of adjoint scalar fields similarly unchanged.  From the known
spectra of the pure gauge theory \cite{mt} and of the fundamental
representation Higgs model \cite{ptw} we expect the spin 0 and spin 2
channels to contain the lowest glueball masses, so in order to carry
out the necessary mixing analysis, we have constructed large bases of
different operator types for these two channels:

\begin{tabbing}
\indent \= $P_d$:\,\, \= \kill
$0^+_+$ channel: \\
\> $R$:   \> $R(x)$ \\
\> $L$:   \> $L_1(x)+L_2(x)$ \\
\> $A$:   \> $\sum_{n=1}^4 R(\theta_n)\left(A^+_{12}(x) + P A^+_{12}(x)\right)$\\
\> $B$:   \> $B_{\perp}(x)$ \\
\> $C$:   \> symmetric combinations of $C^{1\times1},\,C^{2\times2},\,
             C^{1\times2},\,C^{1\times3},\,C^{2\times3}$\\
\> $P$:   \> $P_1^{(L)}(x)+P_2^{(L)}(x)$ \\
\> $P_d$: \> $P_1^{(L)}(x)\cdot P_2^{(L)}(x)$ \\
\> $T$:   \> $\left(P_1^{(L)}(x)\right)^2+\left(P_2^{(L)}(x)\right)^2$ \\
\> \\
$2^+_+$ channel: \\
\> $R$:   \>
$R^{(n)}_{-}(x)\equiv\textstyle\frac{1}{2}
             \left\{\tr\big[\varphi^{(n,1)}(x)
                            \varphi^{(n,1)}(x)\big]
                   -\tr\big[\varphi^{(n,2)}(x)
                            \varphi^{(n,2)}(x)\big] \right\},\quad n>1$ \\
\> $L$:   \> $L_1(x)-L_2(x)$ \\
\> $A$:   \> $\sum_{n=1}^4 {\rm e}^{{\rm i}n\pi}
               R(\theta_n)\left(A^+_{12}(x) + P A^+_{12}(x)\right)$\\
\> $C$:   \> antisymmetric combinations of
             $C^{1\times2},\,C^{1\times3},\,C^{2\times3}$\\
\> $P$:   \> $P_1^{(L)}(x)-P_2^{(L)}(x)$ \\
\> $T$:   \> $\left(P_1^{(L)}(x)\right)^2-\left(P_2^{(L)}(x)\right)^2$ \\
\> \\
\end{tabbing}
In these expressions $R(\theta_n)$ rotates the following brackets
by $\theta_n$ around $x$, and $P$ denotes the parity operation.
The labels
$P,P_d$ refer to single torelon operators, whereas $T$ couples to torelon
pairs in the respective channels.

In principle, similarly extended bases are possible for the other
channels as well.  This requires a lot of memory and computer time,
and furthermore the glueballs in the other channels are anticipated to
be rather heavy
\cite{ptw,mt},
and a mixing analysis would be difficult. The dynamics of mixing is
moreover expected to be independent of the precise quantum number
channel \cite{ptw}, so it should be sufficient to study the $0^+_+$
and $2^+_+$ as detailed examples.  
For the other channels we therefore
limit ourselves to just one or two
operator types $B_{||}$ and/or $A$.  We only give two more explicit
examples; the other combinations follow straightforwardly.
\begin{tabbing}
\indent \= $A$:\,\, \= \kill
$0^{-}_-$ channel: \\
\> $B$:   \> $\sum_{n=1}^4 R(\theta_n) B_{||}(x)$ \\
\> $A$:   \> $\sum_{n=1}^4 R(\theta_n)\left(A^-_{12}(x)-PA^-_{12}(x)\right)$ \\
\> \\

$1^{+}_-$ channel : \\
\> $B$:   \> $\sum_{n=1}^4 {\rm e}^{{\rm i}n\pi/2} R(\theta_n)B_{||}(x)$ \\
\> $A$:  \> $\sum_{n=1}^4 {\rm e}^{{\rm i}n\pi/2} 
                     R(\theta_n)\left(A^-_{12}(x)+PA^-_{12}(x)\right)$\\
\> \\
\end{tabbing}
For all channels we construct operator versions from elementary 
as well as blocked fields at
various blocking levels.
We then take zero momentum sums by averaging over a timeslice,
$\phi(t)= 1/L^2\sum_{x1,x2}\phi(x)$, where $\phi(x)$ denotes a 
generic operator at some blocking level.
 
\subsection{Matrix correlators \label{mc}}

In order to compute the excitation spectra of states with given
quantum numbers, we construct matrix correlators by measuring all
cross correlations between different types of operators at several
blocking levels.  The correlation matrix can be diagonalised
numerically following a variational meth\-od. For a given set of $N$
operators $\{\phi_i\}$ we find the linear combination that minimises the
energy, corresponding to the lightest state.  The first excitation can
be found by applying the same procedure to the subspace
$\{\phi_i\}^{'}$ which is orthogonal to the ground state. This may be
continued to higher states so that we end up with a set of~$N$
eigenstates $\{\Phi_i,\,i=1,\ldots,N\}$ given by
\beq \label{eigen}
\Phi_i(t) =  \sum_{k=1}^N a_{ik}\phi_k(t).
\eeq
The coefficients $a_{ik}$ quantify the overlap of each individual
operator $\phi_k$ used in the simulation onto a particular, approximate
mass eigenstate $\Phi_i$. For a complete basis of operators this
procedure is exact. In practice, the quality of the approximation
clearly depends on the number~$N$ of original operators and their
projection properties. 
The eigenstates already determined are removed from the
basis for the higher excitations, so that the basis for the latter
gets smaller and the corresponding higher states are determined less
reliably.
A more detailed discussion of the calculation and of checks of its stability
can be found in \cite{ptw}, where
the same procedure was applied to the fundamental Higgs model. 

In order to distinguish glueballs and scalar bound states, the most
elaborate calculation is done in the $0^+_+$ and $2^+_+$ channels.
For $0^+_+$ our basis consists of $N=32$ operators $\phi_k\in \{ R, L,
A, B, C, P, T, P_d\}$, for $2^+_+$ we have $N=31$ and $\phi_k\in \{ R,
L, A, C, P, T\}$, each operator considered for at least two blocking
levels. This typically enables us to extract the four or five lowest
lying states in those channels reliably.  In the $0^-_-$, $2^+_-$,
$1^+_-$, $1^-_-$ channels we work with $N=9$ operators $\phi_k\in \{A,
B\}$, and in the remaining channels we have $N=4$ and $\phi_k\in
\{A\}$. Correspondingly, in the smaller bases we can only extract the
ground state and, in some cases, the first excitation reliably.
Finally, in order to compute the string tension we have also
diagonalised a separate basis of Polyakov loop operators at $N=4$
different blocking levels.

\subsection{Simulation and analysis}

The Monte Carlo simulation of the lattice action,
Eq.~(\ref{lattice_action}), is performed using the same algorithm as in
previous work \cite{usad}.  The update of gauge variables employs a
combination of the standard heatbath and over-relaxation algorithms
for SU(2) \cite{ud}, whereas the scalars are updated by a suitable
adaption of the algorithm for fundamental scalars suggested in
\cite{bunk} and employed in \cite{ptw}. In order to account for the
quadratic dependence of the hopping term on the link variables, the
link update for the pure gauge action has to be supplemented by a
Metropolis step.

We define a ``compound" sweep to consist of a combination of one
heatbath and several over-relaxation updates of the gauge and scalar
fields. In this study we performed 5 over-relaxation steps for every
heatbath update.  Measurements are taken after every such compound
sweep. We have gathered between 5 000 and 30 000 measurements,
depending on the lattice sizes which range from $24^3$ for the
coarsest lattice to $48^3$ on the finest one.

All our mass estimates are obtained from measured correlation
functions of operators $\Phi_i$ in the diagonalised bases defined in
the preceding subsection. Correlated fits over some interval $[t_1,t_2]$
are performed
to the ansatz for the asymptotic
behaviour of the correlation function on a finite lattice, for large $T$,
\beq \label{eq:asymp}
\langle\Phi^\dagger_i(t)\Phi_i(0)\rangle
 = A_i\left({\rm e}^{-aM_i\,t} + {\rm e}^{-aM_i(T-t)}\right)\;,
\eeq
where $i$ labels the operator, and $T$ denotes the extent of the
lattice in the time direction.
In cases where the effective masses did not show plateaux long enough to
perform a correlated fit, we used the effective masses for our mass estimates.

Statistical errors are estimated using a jackknife procedure for which 
the individual measurements have been accumulated in bins of $100-250$ sweeps.

\section{Numerical results} 

We have performed simulations at temperatures $T=2T_c$, $T=4T_c$ and
$T=5T_c$ corresponding to three points $\{x,y\}$ in the metastable
confinement phase of the model, as detailed in Table~\ref{tab_params}. 

\begin{table}[t]
\begin{center}
\begin{tabular}{|c|r@{.}l|r@{.}l|l|l|l|}
\hline
\hline
$T$ & \multicolumn{2}{c|}{$x$} & \multicolumn{2}{c|}{$y$} &
$\beta = 9$ & $\beta = 12$ & $\beta = 18$ \\
\hline
$2T_c$  & 0&141 & 0&185 &
        $L^2 \cdot T = 28^3$, $38^3$ & $L^2 \cdot T = 38^3$ & --- \\
$4T_c$  & 0&111 & 0&228 &
        --- & $L^2 \cdot T = 38^3$ & --- \\
$5T_c$  & 0&104 & 0&242 &
        $L^2 \cdot T = 24^3$, $34^3$ & $L^2 \cdot T = 34^3$ & 
        $L^2 \cdot T = 48^3$ \\
\hline
\hline
\end{tabular}
\caption{ \label{tab_params}
{\em The lattice parameters and sizes simulated.}}
\end{center}
\end{table}

The most elaborate test of finite volume effects as well as an explicit
continuum extrapolation for the lowest states are done for $T=5T_c$.
Experience of the pure gauge model
\cite{mt}
suggests that to avoid contamination of the spectrum by light
torelonic states the lattice size should be $L \ge 24$ at $\beta=9$.
We thus simulate three different lattice spacings (the third one being half
of the first) while approximately maintaining
this physical volume ({\it i.e.} $L/\beta = 24/9$,
using the lattice spacing as defined by the bare coupling).
We find all
these couplings to lie in the scaling window of the theory which
permits a continuum extrapolation of the low lying masses. To verify
that finite volume effects are indeed small we also simulate a second,
larger volume at the coarsest lattice spacing.

\begin{table}[t]

\begin{center}
\begin{tabular}{|c|r@{.}l|r@{.}l|r@{.}l|}
\hline
\hline
$T = 2T_c$ &
\multicolumn{4}{|c|}{$\beta=9$}  &
\multicolumn{2}{|c|}{$\beta=12$} \\
\hline
 &\multicolumn{2}{|c|}{$L=38$} &
\multicolumn{2}{|c|}{$L=28$} &
\multicolumn{2}{|c|}{$L=38$} \\
\hline
$aM_P(L)$  &
0&849 (13)   &
0&627 (6)   &
0&462 (4)   
\\
$a\sqrt{\sigma_{\infty}}$ &
0&151 (1) &
0&152 (1)   &
0&112 (1)   
\\
$\sqrt{\sigma_{\infty}}/g_3^2$ &
0&339 (3) &
0&341 (2)   &
0&336 (2)   
\\
\hline
\hline
\end{tabular}
\end{center}

\caption{ \label{2tcc}
{\em Polyakov loop masses and string tensions
at $T=2T_c$.}}
\end{table}

\begin{table}[b]

\begin{center}
\begin{tabular}{|c|r@{.}l|}
\hline
\hline
$T = 4T_c$ &
\multicolumn{2}{|c|}{$\beta=12$} \\
\hline
&\multicolumn{2}{|c|}{$L=38$} \\
\hline
$aM_P(L)$  &
0&476 (4)   
\\
$a\sqrt{\sigma_{\infty}}$ &
0&114 (1)   
\\
$\sqrt{\sigma_{\infty}}/g_3^2$ &
0&342 (2)   
\\
\hline
\hline
\end{tabular}
\end{center}

\caption{ \label{4tcc}
{\em Polyakov loop masses and string tensions
at $T=4T_c$.}}
\end{table}

At $T=2T_c$ we simulate two lattice spacings and physical volumes, and
at $T=4T_c$ just one as we are by now confident that lattice effects are
small. In these cases continuum extrapolation is not possible and
we thus use the masses measured at the smallest lattice spacing as
approximations of the continuum result.

In order to perform continuum extrapolations it is necessary to have
some definition of the lattice spacing with which to scale the
dimensionless lattice masses. The simplest measure comes from the bare
gauge coupling, $a = 4 / (\beta g_3^2)$. Such a formula is liable to
quantum corrections, which in some cases translate as large ${\cal
  O}(a)$ corrections to the na\"{\i}ve scaling of the masses
\cite{moore}.  A non-perturbative definition of the lattice spacing
should in principle avoid this, and in pure gauge theory it is
customary to scale all lattice masses in units of the string tension.
We thus discuss this quantity first.

\subsection{The string tension}

\begin{table}[t]
\begin{center}
\begin{tabular}{|c|r@{.}l|r@{.}l|r@{.}l|r@{.}l|}
\hline
\hline
$T=5T_c$ &
\multicolumn{4}{|c|}{$\beta=9$}  &
\multicolumn{2}{|c|}{$\beta=12$} &
\multicolumn{2}{|c|}{$\beta=18$}  \\
\hline
 &\multicolumn{2}{|c|}{$L=34$} &
\multicolumn{2}{|c|}{$L=24$} &
\multicolumn{2}{|c|}{$L=34$} &
\multicolumn{2}{|c|}{$L=48$} \\
\hline
$aM_P(L)$  &
0&80 (1) &
0&554 (5) &
0&426 (4) &
0&257 (6)
\\
$a\sqrt{\sigma_{\infty}}$ &
0&155 (1) &
0&155 (1) &
0&114 (1) &
0&075 (1)
\\
$\sqrt{\sigma_{\infty}}/g_3^2$ &
0&349 (2) &
0&349 (2) &
0&342 (3) &
0&338 (4)
\\
\hline
\hline
\end{tabular}
\caption{ \label{5tcc}
{\em Polyakov loop masses and string tension
at $T=5T_c$. }}
\end{center}
\end{table}

As explained previously, a Polyakov loop in a spatial plane couples 
to a flux loop state (or torelon)
that winds around the periodic boundaries of the finite volume.
The exponential fall--off of its correlator 
is related to the mass of the flux loop, according to
\beq
\label{string}
  \sum_{\vec{x}}\,\left\langle P_j^{(L)}(x)P_j^{(L)\dag}(0)
                  \right\rangle \simeq e^{-aM_P(L)t},\quad
   aM_P(L)=a^2\sigma_L L.
\eeq
Such a flux loop state can be easily identified through its energy
scaling linearly with the size of the lattice, as seen in the entries
labelled by $P$ in Tables~\ref{2tca},~\ref{4tca} and~\ref{5tca}.  In
contrast to the fundamental Higgs model, there are no fundamental
representation matter fields in the action to screen the colour flux
of static charges leading to a breaking of flux tubes longer than a
screening length
\cite{sb}.  
On the contrary, a flux tube between fundamental charges will persist
to infinite separation as in the pure gauge theory, and a string
tension can be defined in precisely the same way by the slope of the
static potential at infinite separation\footnote{ 
To avoid confusion, we note that this is the string tension in the
$2+1$ dimensional theory, corresponding to the {\it spatial} string tension
in the $3+1$ dimensional theory at finite temperature.}.
Accordingly, periodic flux
loops winding through the boundaries of the lattice will not be
screened, and the coefficient $\sigma_L$ corresponds to the string
tension on a finite volume. An estimate for the string tension in
infinite volume is then provided by the relation
\cite{for85}
\beq
\label{stringinfty}
  a^2\sigma_\infty = a^2\sigma_L+\frac{\pi}{6L^2}.
\eeq
We have diagonalised a basis of Polyakov loop operators and 
the results are summarised in Tables~\ref{2tcc}--\ref{5tcc}.
As we shall see with the glueball masses, the string tension is barely
affected by the presence of the scalar fields. It also shows little
dependence on the scalar parameters $x,y$.

Expressing the string tension in units of the bare coupling, we may
(linearly) extrapolate the results at $T=5T_c$ to the continuum, where
we find $\sqrt{\sigma_\infty}/g_3^2=0.326(7)$, which compares to
$\sqrt{\sigma_\infty}/g_3^2=0.3353(18)$ in pure gauge theory
\cite{mt}. 
That the slope of this extrapolation is small indicates that the
difference between using a tree level definition of the lattice
spacing, and a non--perturbative one is slight, and in most cases any
improvements in using the string tension as opposed to the bare
coupling for setting the scale are outweighed by the additional
statistical errors on the former. 

For this reason, in the following tables we scale and 
quote masses in units of
the bare continuum coupling to allow rapid comparisons of measurements at
different $\beta$.

\subsection{The mass spectrum}

\begin{figure}[t]
\begin{center}
\leavevmode
\epsfysize=250pt
\epsfbox[20 30 620 730]{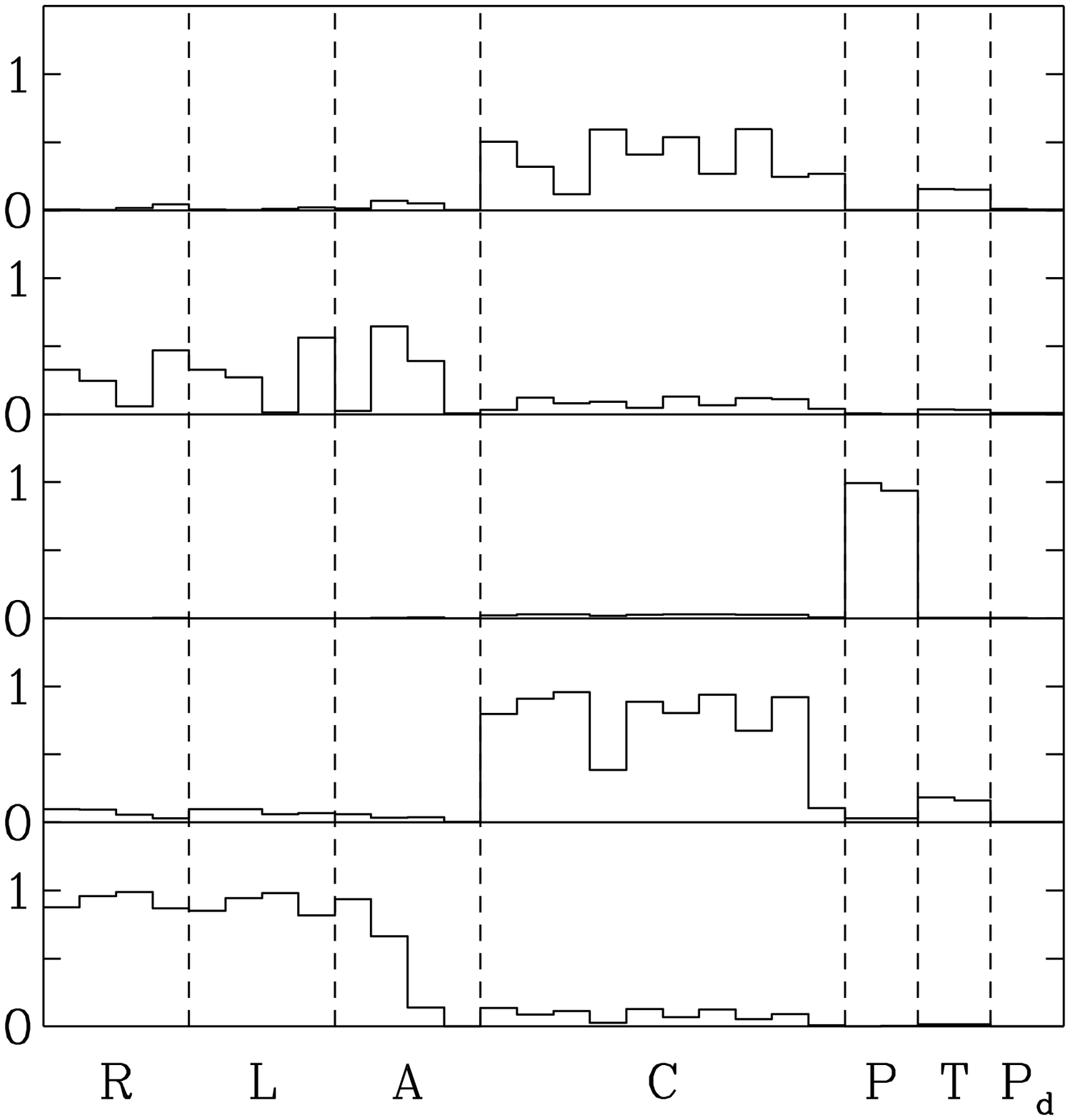}
\leavevmode
\epsfysize=250pt
\epsfbox[20 30 620 730]{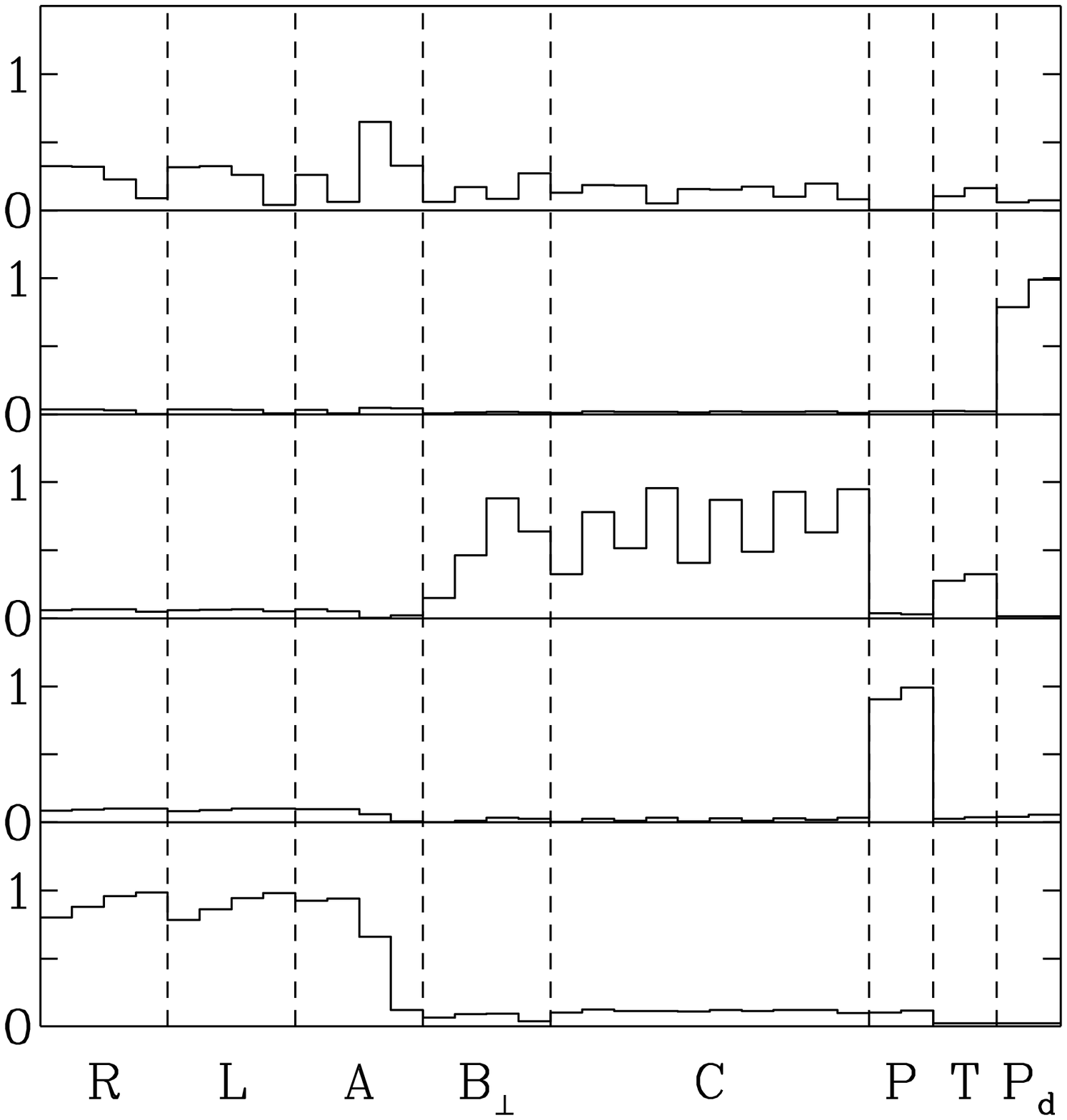}
\end{center}
\vspace{-1.6cm}
\caption[]{\label{op0}
{\it
The coefficients $a_{ik}$ of the operators used in
the simulation for the five lowest $0^+_+$ eigenstates for 28 operators
at $\beta=9$ (left),
and for 32 operators at $\beta=18$ (right).}}
\end{figure}
\begin{table}[t]
\begin{center}
\begin{tabular}{|c|r@{.}lc|r@{.}lc|r@{.}lc|}
\hline
\hline
$T = 2T_c$ &
\multicolumn{6}{|c|}{$\beta=9$}  &
\multicolumn{3}{|c|}{$\beta=12$} \\
\hline
state  &
\multicolumn{3}{|c|}{$L^2\cdot T=38^3$} &
\multicolumn{3}{|c|}{$L^2\cdot T=28^3$} &
\multicolumn{3}{|c|}{$L^2\cdot T=38^3$} \\
\hline
\hline
$0^+_+$  &  \multicolumn{2}{c}{$M/g_3^2$} & Ops.   &
\multicolumn{2}{c}{$M/g_3^2$} & Ops.  &
\multicolumn{2}{c}{$M/g_3^2$} & Ops.   \\
\hline
$\Phi_1$ & 
0&747 (7)   & $R,L,A$     & 
0&743 (7)   & $R,L,A$     & 
0&744 (9)   & $R,L,A$ 
\\
$\Phi_2$ & 
1&629 (27)   & $C,B_\perp$     & 
1&411 (18)   & $P$     & 
1&416 (15)   & $P$ 
\\
$\Phi_3$ & 
2&007 (45)   & $A,(R,L)$     & 
1&613 (23)   & $C,B_\perp$     & 
1&623 (21)   & $C,B_\perp$ 
\\
$\Phi_4$ & 
1&913 (45)   & $P$     & 
1&980 (36)   & $A,(R,L)$     & 
1&998 (27)   & $A,(R,L)$ 
\\
$\Phi_5$ & 
2&22 (6)   & $C,B_\perp$     & 
2&36 (6)   & $C,B_\perp$     & 
2&34 (4)   & $C,(B_\perp)$ 
\\
$\Phi_6$ & 
2&80 (13)  & $A$ &
2&47 (9)   & $P_d$     & 
2&38 (5)   & $P_d$ 
\\
$\Phi_7$ & 
2&94(17) & $C,(B_\perp)$ & 
2&75 (11)   & $A$     & 
2&79 (7)   & $C,B_\perp$ 
\\
$\Phi_8$ & 
3&11 (48)   & $(P)$  &
2&86 (13)   & $C,B_\perp$     & 
2&83 (7)    & $(A)$ 
\\
\hline
\hline
$2^+_+$  &  \multicolumn{2}{c}{$M/g_3^2$} & Ops.   &
\multicolumn{2}{c}{$M/g_3^2$} & Ops.  &
\multicolumn{2}{c}{$M/g_3^2$} & Ops.   \\
\hline
$\Phi_1$ & 
1&94 (5)   & $P$     & 
1&440 (22)   & $P$     & 
1&401 (15)   & $P$ 
\\
$\Phi_2$ & 
1&92 (5)   & $P$     & 
1&454 (18)   & $P$     & 
1&419 (18)   & $P$ 
\\
$\Phi_3$ & 
2&27 (6)   & $R,L,A$     & 
2&24 (5)   & $R,L,A$     & 
2&30 (4)   & $R,L,A$ 
\\
$\Phi_4$ & 
2&61 (11)   & $C$     & 
2&34 (8)   & $(P)$     & 
2&37 (5)   & $(P)$ 
\\
$\Phi_5$ & 
2&60 (29)   & $(P)$     & 
2&70 (12)   & $C$     & 
2&65 (5)   & $C$ 
\\
$\Phi_6$ & 
3&15 (19)   & $C$     & 
3&05 (15)   & $C,T$     & 
2&98 (8)   & $C,T$ 
\\
\hline
\hline
\end{tabular}
\caption{ \label{2tca}
{\em Mass estimates and dominant operator
contributions in the $0^+_+$ and $2^+_+$ channels at 
$T=2T_c$. The dominant operator types contributing 
are denoted with (without) parentheses if $a_{ik}<(>)0.5$.}}
\end{center}
\end{table}

\begin{table}

\begin{center}
\begin{tabular}{|l|r@{.}l|r@{.}l|r@{.}l|r@{.}l|}
\hline
\hline
$T = 2 T_c$ &
\multicolumn{4}{|c|}{$\beta=9$}  &
\multicolumn{2}{|c|}{$\beta=12$} \\
\hline
state  &
\multicolumn{2}{|c|}{$L=38$} &
\multicolumn{2}{|c|}{$L=28$} &
\multicolumn{2}{|c|}{$L=38$} \\
\hline
$0^-_+$  &
\multicolumn{2}{c|}{---} &
\multicolumn{2}{c|}{---} &
3&24 (12) 
\\
\hline
$0^-_-$  & 
1&229 (14)   & 
1&220 (14)   & 
1&118 (12)   
\\
$0^{-*}_-$  & 
2&14 (6)   & 
2&14 (6)   & 
2&16 (4)   
\\
\hline
$1^+_+$ & 
\multicolumn{2}{c|}{---}   & 
\multicolumn{2}{c|}{---}   & 
3&04 (12)   
\\
\hline
$1^-_+$  & 
3&00 (5)   & 
3&01 (5)   & 
2&93 (11)   
\\
\hline
$1^+_-$ & 
2&02 (2)   & 
2&01 (2)   & 
1&99 (3)   
\\
\hline
$1^-_-$  & 
1&95 (5)   & 
1&87 (8)   & 
1&92 (5)   
\\
\hline
$2^+_-$  & 
2&41 (9)   & 
2&57 (9)   & 
2&52 (6)   
\\
\hline
$2^-_+$  & 
2&35 (3)   & 
2&36 (6)   & 
2&31 (5)   
\\
\hline
\hline
\end{tabular}
\end{center}

\caption{ \label{2tcb}
{\em Masses $M/g_3^2$ for the
other channels at $T=2T_c$.}}
\end{table}

\begin{table}
\begin{center}
\begin{tabular}{|c|r@{.}lc|r@{.}lc|r@{.}lc|}
\hline
\hline
$T = 4T_c$ &
\multicolumn{3}{|c|}{$\beta=12$} \\
\hline
state  &
\multicolumn{3}{|c|}{$L^2\cdot T=38^3$} \\
\hline
\hline
$0^+_+$  &  
\multicolumn{2}{c}{$M/g_3^2$} & Ops.   \\
\hline
$\Phi_1$ & 
0&945 (9)   & $R,L,A$ 
\\
$\Phi_2$ & 
1&425 (15)   & $P$ 
\\
$\Phi_3$ & 
1&626 (18)   & $C,B_\perp$ 
\\
$\Phi_4$ & 
2&151 (30)   & $A,(R,L)$ 
\\
$\Phi_5$ & 
2&33 (4)   & $C, B_\perp$ 
\\
$\Phi_6$ & 
2&33 (4)   & $(P)$ 
\\
$\Phi_7$ & 
2&43 (5)   & $P_d$ 
\\
$\Phi_8$ & 
2&84 (7)    & $C$ 
\\
\hline
\hline
$2^+_+$  &  
\multicolumn{2}{c}{$M/g_3^2$} & Ops.   \\
\hline
$\Phi_1$ & 
1&428 (21)   & $P$ 
\\
$\Phi_2$ & 
1&452 (18)   & $P$ 
\\
$\Phi_3$ & 
2&42 (4)   & $R,L,A$ 
\\
$\Phi_4$ & 
2&34 (5)   & $(P)$ 
\\
$\Phi_5$ & 
2&63 (6)   & $C$ 
\\
$\Phi_6$ & 
3&02 (8)   & $C,T$ 
\\
\hline
\hline
\end{tabular}
\caption{ \label{4tca}
{\em Mass estimates and dominant operator
contributions in the $0^+_+$ and $2^+_+$ channels at 
$T=4T_c$. The dominant operator types contributing 
are denoted with (without) parentheses if $a_{ik}<(>)0.5$}}
\end{center}
\end{table}

\begin{table}

\begin{center}
\begin{tabular}{|l|r@{.}l|}
\hline
\hline
$T = 4 T_c$ &
\multicolumn{2}{|c|}{$\beta=12$} \\
\hline
state  &
\multicolumn{2}{|c|}{$L=38$} \\
\hline
$0^-_+$  &
3&48 (12) 
\\
\hline
$0^-_-$  & 
1&323 (12)   
\\
$0^{-*}_-$  & 
2&30 (4)   
\\
\hline
$1^+_+$ & 
3&47 (24)   
\\
\hline
$1^-_+$  & 
3&05 (11)   
\\
\hline
$1^+_-$ & 
2&00 (3)   
\\
\hline
$1^-_-$  & 
2&05 (4)   
\\
\hline
$2^+_-$  & 
2&57 (5)   
\\
\hline
$2^-_+$  & 
2&45 (5)   
\\
\hline
\hline
\end{tabular}
\end{center}

\caption{ \label{4tcb}
{\em Masses $M/g_3^2$ for the
other channels at $T=4T_c$.}}
\end{table}

\begin{table}
\begin{center}
\begin{tabular}{|c|r@{.}lc|r@{.}lc|}
\hline
\hline
$T=5T_c$ &
\multicolumn{6}{|c|}{$\beta=9$}  \\
\hline
state  &
\multicolumn{3}{|c|}{$L^2\cdot T=34^3$} &
\multicolumn{3}{|c|}{$L^2\cdot T=24^3$} \\
\hline
\hline
$0^+_+$  &  \multicolumn{2}{c}{$M/g_3^2$} & Ops.   &
\multicolumn{2}{c}{$M/g_3^2$} & Ops. \\
\hline
$\Phi_1$ & 
1&010 (9)  & $R,L,A$ & 
1&010 (12) & $R,L,A$ 
\\
$\Phi_2$ & 
1&61 (5)   & $C$ &
1&24 (3)   & $P$   
\\
$\Phi_3$ & 
1&82 (7) & $P$     & 
1&64 (3) & $C,B_{\perp}$
\\
$\Phi_4$ & 
2&19 (14)  & $L,A$   &
2&16 (11)  & $L,A$   
\\
$\Phi_5$ & 
2&29 (16)  & $C$ &
2&05 (13)  & $P_d$  
\\
$\Phi_6$ & 
2&73 (15) & $(P)$      & 
2&36 (18) & $C,B_{\perp}$ 
\\
\hline
\hline
$2^+_+$  &  \multicolumn{2}{c}{$M/g_3^2$} & Ops.   &
\multicolumn{2}{c}{$M/g_3^2$} & Ops. \\
\hline
$\Phi_1$ & 
1&795 (32) & $P$ & 
1&215 (25) & $P$ 
\\
$\Phi_2$ & 
1&798 (32) & $P$ & 
1&253 (25) & $P$ 
\\
$\Phi_3$ & 
2&35 (8) & $R,L,A$ & 
2&25 (10) & $(P)$ 
\\
$\Phi_4$ & 
2&68 (11) & $C$ & 
2&45 (15) & $R,L,A$ 
\\
$\Phi_5$ & 
2&67 (11) & $(P)$ & 
2&57 (11) & $C,T$ 
\\
$\Phi_6$ & 
2&87 (18) & $C$ &
2&86 (11) & $C,T$ 
\\
$\Phi_7$ & 
3&34 (24) & $R,L,A$ & 
3&17 (17) & $R,L,A$ 
\\
\hline
\multicolumn{7}{c}{\mbox{}} \\
\hline
$T=5T_c$ &
\multicolumn{3}{|c|}{$\beta=12$} &
\multicolumn{3}{|c|}{$\beta=18$}  \\
\hline
state  &
\multicolumn{3}{|c|}{$L^2\cdot T=34^3$} &
\multicolumn{3}{|c|}{$L^2\cdot T=48^3$} \\
\hline
\hline
$0^+_+$  &  \multicolumn{2}{c}{$M/g_3^2$} & Ops.   &
\multicolumn{2}{c}{$M/g_3^2$} & Ops.   \\
\hline
$\Phi_1$ & 
1&005 (12) & $R,L,A$ & 
1&008 (18) & $R,L,A$ 
\\
$\Phi_2$ & 
1&26 (3)   & $P$ & 
1&152 (27) & $P$  
\\
$\Phi_3$ & 
1&63 (3) & $C,B_{\perp}$ & 
1&60 (5) & $C,B_{\perp}$ 
\\
$\Phi_4$ & 
2&19 (8)   & $A$ & 
1&91 (5)   & $P_d$  
\\
$\Phi_5$ & 
2&33 (8)   & $C,B_{\perp},(T)$ & 
2&19 (7)   & $A$ 
\\
$\Phi_6$ & 
2&25 (8)  & $P_d$ & 
2&29 (9)  & $C,T$ 
\\
\hline
\hline
$2^+_+$  &  \multicolumn{2}{c}{$M/g_3^2$} & Ops.   &
\multicolumn{2}{c}{$M/g_3^2$} & Ops.   \\
\hline
$\Phi_1$ & 
1&266 (21) & $P$ & 
1&143 (27) & $P$ 
\\
$\Phi_2$ & 
1&274 (24) & $P$ & 
1&161 (27) & $P$ 
\\
$\Phi_3$ & 
2&32 (4)  & $(P)$ & 
2&38 (5)  & $(P)$ 
\\
$\Phi_4$ & 
2&43 (5) & $R,L,A$ & 
2&51 (7) & $C,T$ 
\\
$\Phi_5$ & 
2&61 (5)  & $C,(T)$ & 
2&45 (8)  & $L,A$ 
\\
$\Phi_6$ & 
2&78 (6) & $C,T$ & 
2&72 (6) & $C$ 
\\
$\Phi_7$ & 
3&23 (11) & $L$ & 
3&17 (10)  & $A$ 
\\
\hline
\hline
\end{tabular}
\caption{ \label{5tca}
{\em Mass estimates and dominant operator
contributions in the $0^+_+$ and $2^+_+$ channels at 
$T=5T_c$. The dominant operator types contributing 
are denoted with (without) parentheses if $a_{ik}<(>)0.5$.}}
\end{center}
\end{table}

\begin{table}
\begin{center}
\begin{tabular}{|l|r@{.}l|r@{.}l|r@{.}l|r@{.}l|}
\hline
\hline
$T=5T_c$&
\multicolumn{4}{|c|}{$\beta=9$}  &
\multicolumn{2}{|c|}{$\beta=12$} &
\multicolumn{2}{|c|}{$\beta=18$}  \\
\hline
state  &
\multicolumn{2}{|c|}{$L^2\cdot T=34^3$} &
\multicolumn{2}{|c|}{$L^2\cdot T=24^3$} &
\multicolumn{2}{|c|}{$L^2\cdot T=34^3$} &
\multicolumn{2}{|c|}{$L^2\cdot T=48^3$} \\
\hline
$0^-_+$  &
\multicolumn{2}{c|}{---} &
\multicolumn{2}{c|}{---} &
3&39 (15)  &
\multicolumn{2}{c|}{---}
\\
\hline
$0^-_-$  &
1&352 (27) &
1&368 (28) &
1&342 (33) &
1&340 (30)
\\
$0^{-*}_-$  &
2&34 (8)  &
2&28 (6) &
2&33 (8) &
2&28 (5)
\\
\hline
$1^+_+$  &
3&01 (37) &
3&65 (37) &
3&52 (37)  &
\multicolumn{2}{c|}{---}
\\
\hline
$1^-_+$  &
3&02 (26) &
3&16 (20) &
3&20 (20)  &
\multicolumn{2}{c|}{---}
\\
\hline
$1^+_-$  &
2&13 (3) &
2&04 (5) &
2&06 (6)  &
\multicolumn{2}{c|}{---}
\\
\hline
$1^-_-$  &
2&02 (6) &
2&09 (5) &
2&06 (5)  &
\multicolumn{2}{c|}{---}
\\
\hline
$2^+_-$  &
2&69 (4) &
2&60 (10) &
2&64 (9)  &
\multicolumn{2}{c|}{---}
\\
\hline
$2^-_+$  &
2&52 (10) &
2&60 (9) &
2&54 (10)  &
\multicolumn{2}{c|}{---}
\\
\hline
\hline
\end{tabular}
\caption{ \label{5tcb}
{\em Masses $M/g_3^2$ for the
other channels at $T=5T_c$.}}
\end{center}
\end{table}

A detailed account of the mass estimates for the lowest states
computed on the various lattices is given in
Tables~\ref{2tca}--\ref{5tcb}, together with the dominant operator
content of a given mass eigenstate for the channels with the largest
operator bases.

Let us discuss in detail only the $T=5T_c$ parameter set, the
procedure being entirely analogous for the others. As an example
for the identification of the operator content of a mass eigenstate,
we display the coefficients $a_{ik}$ ({\it cf.} Eq.~(\ref{comp})) of
the five lowest mass eigenstates in the $0^+_+$-channel in
Fig.~\ref{op0} for our largest and smallest lattice spacing. The
operator contributions listed in Table~\ref{5tca} are those which have
coefficients $a_{ik}$ larger than 0.5. If the state has weak overlap
with all operators of our basis and the dominant coefficients are
smaller than 0.5, we indicate this by parentheses in the table. The
ordering of states $\Phi_i$ was obtained during the diagonalisation
procedure from the effective mass of the corresponding correlator on
the first timeslice.

As in the confinement phase of the fundamental representation Higgs
model~\cite{ptw}, we find here a dense spectrum of bound states also.
In \cite{ptw} it was found that the Yang-Mills glueball spectrum is
repeated almost identically in the model with fundamental scalar
fields, with additional bound states of scalars, and little mixing
between them.  An interesting question is whether the situation is
similar in the case with adjoint scalar fields.  Fig.~\ref{op0} (left)
shows that this is indeed so. The ground state is clearly
dominated by scalar operators, whereas the first as well as the fourth
excited state have almost exclusively gluonic contributions and hence
are identified as glueballs.  Fig.~\ref{op0} (right) is the same for our
finest lattice, but now with four blocked versions of the operator
$B_{\perp}$ ({\it cf.}  Eq.~(\ref{bop})) included in addition to the
operators used for the left plot. At first sight this appears to
violate the ``non-mixing'' between scalar and gluonic states as
observed with the smaller basis.  Upon expanding and taking the trace
in Eq.~(\ref{bop}), however, one readily sees that the operator is in
fact a sum of terms of the form $B_{\perp}\sim \tr(U^2) + B^2_{||} +
\tr(\varphi^2) B^2_{||}$.  Of these, the first term mixes with our other
loop operators and has projection onto the glueball states since
($\tr(U^2)$ has the same representation content as $\tr(U)$), whereas
the other terms project onto higher excitations (they are of higher
dimension) and thus die out asymptotically.  Hence, the operator
$B_{\perp}$ asymptotically behaves as a $C$-type operator which mainly
projects onto glueballs, but very little onto scalar states, as is
corroborated in Fig.~\ref{op0}. This figure further shows that the
decoupling of the gluonic sector is stable under variations of the
lattice spacing by a factor of two.  As we shall see, with the lattice
spacings considered we are quite close to the continuum, and the
observed effect should persist in the continuum limit.
This picture is repeated in the $2^+_+$ channel.  

The spectrum of higher excitations becomes increasingly dense with
increasing energy. For the $0^+_+$ and $2^+_+$ channels, the
eigenstates beyond $\Phi_6$ become increasingly difficult to identify,
and their total overlap with the operators in the basis becomes
smaller.  Further, they are closer in energy and within our
statistical errors appear often degenerate. An unambiguous resolution
of these states would require still larger bases as well as more
statistics. We have therefore chosen not to quote them as reliable
results.

Another word of caution concerns the states $0^-_+$, $1^+_+$ and
$1^-_+$.  As described in Sec.~\ref{mc}, the bases for these states
consist exclusively of four blocked versions of the operator type $A$.
It is conceivable that these operator sets have very poor projection
onto some of the sought ground states, leading to an overestimate of
their masses.  Indeed, such an effect was encountered in the $1^P_-$
sectors, as is apparent from comparing the two different parity
states, which should be degenerate.  This problem can only be solved
with a larger basis including other operator types. For $2^-_+$ the
basis equally consists of four operators of type $A$. In this case,
however, we observe parity doubling with the ground state in the
$2^+_+$ channel (see below), and may thus be confident to have good
projection on the $2^-_+$ ground state.

At our largest lattice spacing, $\beta=9$, we have performed an
explicit check for finite volume effects. Comparison of the
corresponding columns in Table~\ref{5tca} shows that the masses for
those states that do not couple to winding operators ($P,P_d,T$) are
statistically compatible on $L^2\cdot T=24^3$ and $L^2\cdot T=34^3$.
It appears that already on the smaller volume most of our results for
the spectrum are free of finite size effects, as was the case in the
pure gauge theory.  Of the operator eigenvectors projecting onto
winding operators, those purely of type $P$ or $P_d$ may be discounted
as mere finite volume artefacts, they disappear from the spectrum in
the infinite volume limit.  More difficult are those states that
couple both to non-winding, $C$ or $B_\perp$, and to winding
operators, $T$, such as the $2^+_+$ states $\Phi_5$ and $\Phi_6$ whose
masses are likely to suffer most from finite volume effects.  At
$\beta=9$, where we have two volumes, we can resort to the larger
lattice where the states are decoupled from the torelons, and trust
that this number is as close to the infinite volume limit as the other
states.  On the finer lattices, however, we have only one volume, and
hence it is likely that the corresponding masses are affected by the
admixture of torelons to these states.

As we have discussed, one important signal of finite volume and
discretisation effects is the loss of parity doubling in the spectrum.
On all our lattices we find within statistical errors a degeneracy of
the $J^+_R$ and $J^-_R$ states not only for $J=1$ (where it is
expected even on the lattice) but also for $J=2$. 
In the latter case, the relevant comparison is between non--torelonic
states that do not couple to winding operators (and thus remain of
finite mass in the infinite volume extrapolation).

We have already commented that in extrapolating to the continuum limit
there is a choice in the method of defining the lattice spacing. There
is also some freedom in the functional form of the extrapolation. From
the weak coupling expansion of the lattice action, the lowest order
corrections to the continuum limit mass should be linear in the
lattice spacing. Thus if we are sufficiently close to the continuum
limit, a straight line fit to the data can be attempted. 

\begin{figure}[tbp]
%


\begin{center}
\leavevmode
\epsfysize=280pt
\epsfbox[20 30 620 730]{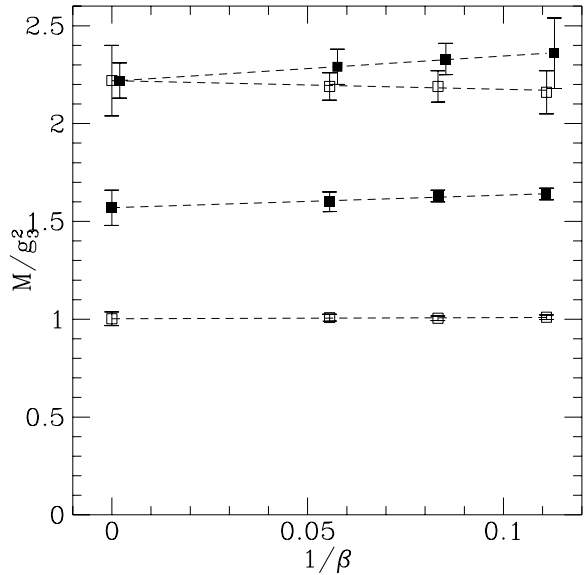}
\end{center}
\vspace*{-2.0cm}
\caption[]{\label{cont0}
{\it
The approach to the continuum of the 
four lowest $0^+_+$ states at $T=5T_c$. 
Filled symbols denote glueballs, open symbols
bound states of scalars.}}
\end{figure}

We have performed linear fits for the low lying $0^+_+$ and $2^+_+$
states, using both tree level and non-perturbative definitions of the
lattice spacing. All fits give acceptable $\chi^2/{\rm d.o.f.} < 1$.
In Fig.~\ref{cont0} we show such fits using units of the bare
coupling, and it is such a fit that we use for the ground state in the $2^+_+$
channel.  
Given the very large errors on the gradient for linear fits
using the string tension to set the scale, we have insufficient data
to decide if the latter, non-perturbative definition of the lattice
spacing improves the scaling behaviour.

The results for $0^+_+$ and $2^+_+$ are given in Table~\ref{res02}.
The $2^+_+$ pure glueball couples to torelons on our finest lattice,
so that its mass value there could not be used for an extrapolation.
We quoted the value for $\beta=12$ instead.  We also compare with the
corresponding glueball states in the pure gauge theory and find their
masses to deviate at the percent level at most from those measured in
the Higgs model.  We conclude that we have a situation analogous to
the fundamental Higgs model, with the pure gauge glueball spectrum
being repeated, with additional bound states of adjoint scalars and
little mixing between them.

As comparison of Tables~\ref{res02} and~\ref{5tca} elucidates, the
difference between the masses extrapolated to the continuum and those
computed at finite lattice spacing are very small, of the order of a
few percent for the lowest states, and mostly covered by the
statistical errors of the values on the finest lattice.  We therefore
quote the latter for the other channels as well as our other parameter
sets, noting, however, that for the heavier states this may well
amount to an estimated 5\% systematic error.

\begin{table}
\begin{center}
\begin{tabular}{||r@{.}lr@{.}l|r@{.}l||r@{.}lr@{.}l|r@{.}l||}
\hline
\hline
\multicolumn{6}{||c||}{$0^+_+$ channel} &
\multicolumn{6}{c||}{$2^+_+$ channel} \\
\hline
\multicolumn{4}{||c|}{gauge-Higgs} & \multicolumn{2}{c||}{pure gauge} &
\multicolumn{4}{c|}{gauge-Higgs} & \multicolumn{2}{c||}{pure gauge}
\\
\hline
\multicolumn{2}{||c}{scalar} & \multicolumn{2}{c|}{glueball} &
\multicolumn{2}{c||}{glueball} &
\multicolumn{2}{c}{scalar} & \multicolumn{2}{c|}{glueball} &
\multicolumn{2}{c||}{glueball} \\
\hline
1&002 (35)     &  \mcemptyr    &  \mcemptyrr  &
2&46 (21)    &  \mcemptyr    &  \mcemptyrr  \\
\mcemptyll  &  1&57 (9)     &  1&58 (1)  &
\mcempty   &  2&61 (5)     &  2&62 (5) \\
2&22 (18)    &  \mcemptyr    &  \mcemptyrr    &
\mcempty   &  \mcemptyr      &  \mcemptyrr    \\
\mcemptyll  &  2&22 (25)     &  2&29 (3)    &
\mcempty   &  \mcemptyr      &  \mcemptyrr  \\
\hline
\hline
\end{tabular}
\caption{ \label{res02}
{\em Final mass estimates using the continuum extrapolation in the
bare coupling of our data at $T=5T_c$. Our data for
the glueball are compared with the results obtained in the pure gauge
theory \cite{mt}.}}
\end{center}
\end{table}
\section{Discussion}

\begin{figure}[tbp]
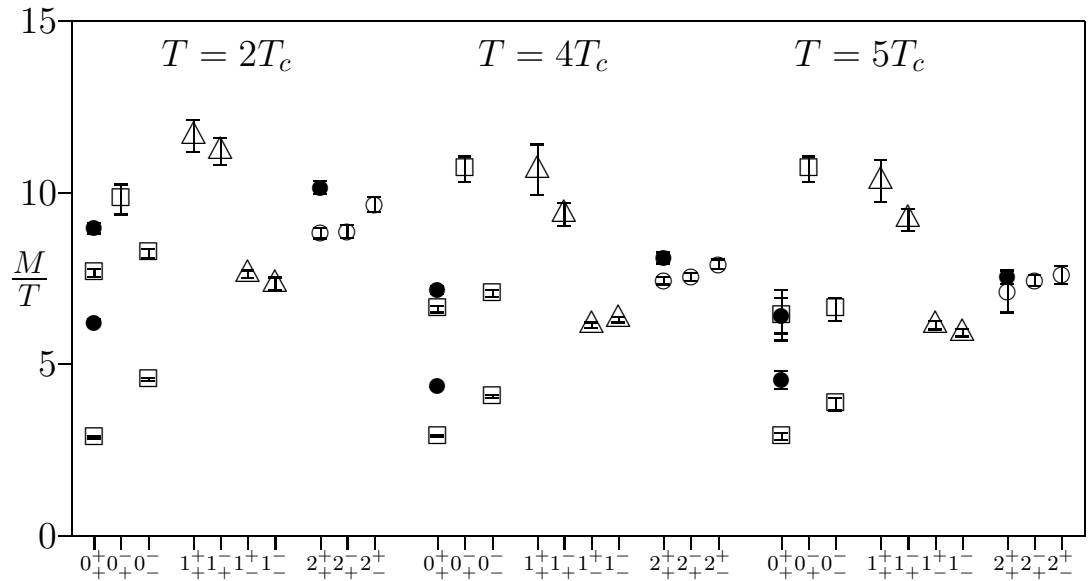


\hspace{-3.5em}
\begin{minipage}{6.5in}
\include{m_over_t}
\end{minipage}

\caption[]{\label{spec}
{\it
The spectrum of the lowest inverse static correlation lengths 
at each temperature. Filled symbols denote glueballs, open symbols
bound states of scalars.}}
\end{figure}

With the numerical results for various correlation functions at hand,
let us now discuss their implications for the spectrum
of static correlation functions
in four-dimensional SU(2) Yang-Mills theory at finite temperature.
In order to express the
masses entirely in units of temperature, we have to make use of the
perturbative expression for the gauge coupling at that temperature scale,
Eq.~(\ref{3d}). Using this expression we have $g_3^2(2T_c)\approx 3.83 T$,
$g_3^2(2T_c)\approx 3.07 T$ and $g_3^2(5T_c)\approx 2.89 T$.
The spectrum in these units is shown in
Fig.~\ref{spec}.

\subsection{Comparison with four dimensions}

The first important question concerns the comparison with a
four-dimensional simulation, in order to assess the quality of
dimensional reduction in the temperature range under study. A detailed
analysis of the low lying modes for SU(2) gauge theory has recently
been presented \cite{dg} for the temperature range $2T_c<T<4T_c$.
Only the ground states in each quantum number channel are so
far available in four dimensions, however.

To facilitate the comparison we briefly summarise the situation in
four dimensions. There the lattice symmetry group is $D_{4h} \otimes
Z(2)_{R_{\bar{z}}}$ where $D_{4h}$ is the point group of the
discretised $2+1$ dimensional space-time slice used to create
screening state operators, and $R_{\bar{z}}$ is the Euclidian time
reversal operator in the notation of \cite{ay}.  It acts on the time
component of the gauge field as $R_{\bar{z}} A_0^a\to -A_0^a$.
Dimensional reduction converts the $A_0$ component of the gauge field
into an adjoint scalar field, and the scalar reflection symmetry is
the ``daughter" of Euclidian time reflection in the original theory.

As the temperature is increased in the four-dimensional theory, the
timelike direction of the slice becomes squeezed, to the point where
the slice becomes quasi--two-dimensional, and the effective symmetry
group becomes $C^4_v$. At this point the spectrum of screening masses
should conform to a pattern predicted by this latter symmetry group,
as has indeed been observed
\cite{dg}.
If the perturbative matching of finite temperature SU(2)
pure gauge theory to
the SU(2) adjoint Higgs model in three dimensions is correct, the
screening masses should furthermore match those observed in this study. 
Hence
we expect to see the $3+1$ dimensional $D_{4h}$ states $A^+_1$ and $A^-_2$
converging on the $0^+$ mass we measure (the quantum numbers $R_{\bar{z}}$,
$R$ are suppressed here), and the $A^-_1$ and $A^+_2$ on the $0^-$.
The $E$ states should agree with the $J=1$. The $B^+_1$ and $B^-_2$
should agree with the $2^+$, and $B^-_1$ and $B^+_2$ should have a
common mass with the $2^-$. 
For a more detailed account of this symmetry breaking pattern we
refer to \cite{dg}.

In \cite{dg,dg2} and the above paragraph discussing the mapping of
states, the behaviour under Euclidian time reflection $R_{\bar{z}}$
has not been specified.  This leaves room for potential ambiguities in
comparing the states.  For our comparison, we assume that the lowest
states have been extracted in \cite{dg}, without explicit
specification of their $R_{\bar{z}}$-symmetry.  Correspondingly, for
the same spin and parity we compare with the lighter of our
$R$-states. Further, because of parity doubling, $B_1^-$ has to be
degenerate with $B_1^+$ in the continuum limit. In \cite{dg2}, $B^+_1$
still exhibits finite size effects, so that we take $B_1^-$ instead to
compare with our $2^+_+$ state. The four-dimensional data are taken
from Table 4 in \cite{dg} and Table 2 in \cite{dg2}.

\begin{figure}[tbp]

\begin{center}
\leavevmode
\epsfysize=280pt
\epsfbox[20 30 620 730]{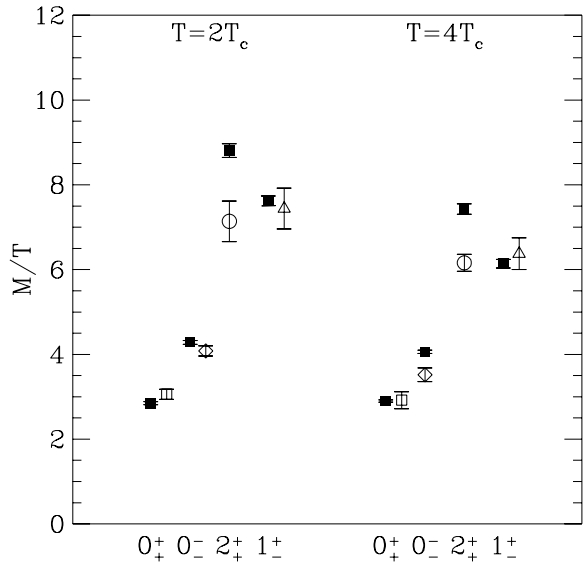}
\vspace{-2.0cm}
\end{center}
\caption{ \label{comp}
{\em Comparison of screening masses as calculated in four dimensions 
directly \cite{dg}
(open symbols)
and from the reduced three-dimensional theory used here (full symbols). 
In the point group 
notation of \cite{dg}, the open squares denote $A_1^+$, 
diamonds $A_1^-$, circles $B_1^-$ and triangles $E^+$.}}
\end{figure}

The comparison of screening masses after these identifications is
shown in Fig.~\ref{comp}.  Comparing the temperature dependence of the
screening masses, we find agreement for the qualitative behaviour of
the lowest states in all channels. Their temperature dependence is
weak, with the $0^+_+$ slightly rising whereas all other ground states
drop with $T$.  In our simulation we find in addition that the
temperature dependence becomes stronger for the higher excitations,
{\it cf.}~Fig.~\ref{spec}. But even on a quantitative level, we find a
remarkable agreement for the low lying states. This could not have
been anticipated given that the temperature is rather close to the
phase transition, where we know the effective theory description has
to break down.  The only notable disagreement between the measurement
in the full and the reduced theory is for the lightest spin 2 screening
mass.  But in comparing the heavier screening masses we have to remind
ourselves that the effective theory is obtained by integrating out the
non-zero Matsubara modes, whose smallest is of the order of $\sim 2\pi
T$.  The lowest spin 2 screening mass, on the other hand, is already
larger than $\sim 6T$ for the temperatures considered, and hence
belongs to an energy region where the effective theory loses its
validity.  In this r\'egime the non-local terms and terms of higher
dimension which have been discarded in the construction of the
effective theory \cite{ad} are no longer suppressed, and disagreement
for screening masses of that order and higher is in fact expected.

We conclude that the three-dimensional
SU(2) adjoint Higgs model in its metastable symmetric phase is the
correct effective theory to describe the static correlation functions
in a four-dimensional SU(2) pure gauge theory at finite temperature.
The effective theory gives the correct qualitative
picture, and for the screening masses in the expected range of validity
of the effective theory even quantitative
results at temperatures as low as $T=2T_c$.

\subsection{The Debye mass}

Of particular interest for non-Abelian plasma physics is the Debye
screening mass, which may be expanded as
\begin{equation}
m_D = m_D^{LO}+{Ng_3^2\over4\pi}\ln{m_3\over g_3^2}
+c_N g_3^2 + {\cal O}(g^3T)\label{md4d}.
\end{equation}
The leading order result is known perturbatively \cite{lo} to be
\beq
m_D^{LO}=m_3=\left(\frac{N}{3}+\frac{N_f}{6}\right)^{1/2}gT.
\eeq
At next-to leading order $\sim g^2T$, one can extract perturbatively
a logarithm \cite{rebhan}, but $c_N$
is entirely non-perturbative
and has to be evaluated by lattice simulations.  A gauge
invariant, non-perturbative definition for $m_D$ has been given in
\cite{ay} and is based on Euclidian time reflection,
$R_{\bar{z}}$.
According to the definition in \cite{ay}, the Debye mass $m_D$
corresponds to the mass of the lightest state odd under this
transformation, {\it i.e.} the lightest $R=-1$ state.  According to
our measurements this is the $0^-_-$ ground state, as obtained from
the operator $B_{||}$. As discussed in the previous section, our
measurement is entirely consistent with the four-dimensional one, see
Fig. \ref{comp}.

On the other hand, $c_2$ has been determined directly in recent
simulations \cite{lp1,lp2} after integrating out $A_0$ following a
prescription in \cite{ay}. There the value $c_2=1.14(4)$ was found for
the leading correction to the perturbative result%
\footnote{We remark that the operator $B_{||}$ has been simulated
  previously to determine the Debye mass \cite{mD}.  By fitting
  Eq.~(\ref{md4d}) to the temperature dependence of the mass of
  $B_{||}$, these authors find the term ${\cal O}(g^3T)$ to be
  consistent with zero, and upon subtracting the perturbative piece
  $m_D^{LO}$ determine the next-to-leading order correction to be
  $c_2=1.58(20)$. We consider this likely to be an overestimate, as it is
  two standard deviations above the results of
  \cite{lp1,lp2}.  Adding the perturbative contribution and no other 
  corrections, the resulting $m_D$ likewise deviates
  from the four-dimensional result of \cite{dg} and our present
  measurements.}.

Taking these data together, we are now in a position to assess
the size of the various contributions in the expansion Eq.~(\ref{md4d}).
These are summarised in Table~\ref{deb}.

The emerging picture is quite interesting: the Debye mass is entirely
non-perturb\-ative, with its leading order value being much smaller
than the $g^2T$ correction up to temperatures of the order of $T\sim
10^7T_c$ \cite{lp2}. On the other hand, the ${\cal O}(g^3T)$
corrections are only 20\% at temperatures as low as $T=2T_c$,
decreasing rapidly to 10\% at $T=5T_c$.

We thus have to conclude that the scale dominating the physics of
electric screening is the scale of the three-dimensional effective
theory $g_3^2$, or equivalently, the soft modes~$\sim g^2T$ of the
original finite temperature theory, contrary to the na\"{\i}ve
expectation~$\sim gT$.  The contributions on this scale are entirely
non-perturbative.  Contributions of higher powers in $g$, on the other
hand, can be viewed as small perturbations.

\begin{table}
\begin{center}
\begin{tabular}{|c|r@{.}l|r@{.}l|r@{.}l|r@{.}l|}
\hline
\hline
 &\multicolumn{2}{|c|}{$m_D/g_3^2$} &
\multicolumn{2}{|c|}{$(1+2)/g_3^2$} &
\multicolumn{2}{|c|}{$c_2$} &
\multicolumn{2}{|c|}{${\cal O}(g^3T)/g_3^2$} \\
\hline
$T=2T_c$  & 1&12 (2) & 0&296 & 1&14(4) & -0&32(6) \\
$T=4T_c$  & 1&32 (2) & 0&360 & 1&14(4) & -0&18(6) \\
$T=5T_c$  & 1&34 (3) & 0&379 & 1&14(4) & -0&18(7) \\
\hline
\hline
\end{tabular}
\caption{ \label{deb}
{\em The different contributions to the Debye mass, Eq.~(\ref{md4d}),
where (1+2) is the sum of the first two terms.}}
\end{center}
\end{table}

\section{Conclusions}

We have performed a detailed simulation of the spectrum of the 
SU(2) adjoint Higgs model in $2+1$ dimensions at various points in its
metastable confinement phase. The choice of the parameter values
was motivated by the connection with four-dimensional SU(2) pure
gauge theory at finite temperature, for which the adjoint Higgs model 
emerges as an effective theory for the static modes after perturbative
dimensional reduction. Our results are relevant both for 
the study of confinement in three-dimensional gauge theories, as
well as for the question whether dimensional reduction is applicable to 
SU(2) pure gauge theory, and eventually to QCD.

Regarding three-dimensional gauge theories,
we found a dense spectrum of bound states, which consists of an almost
unchanged replica of the glueball spectrum of pure gauge theory, 
to which additional bound states of adjoint scalars are added. 
The glueball and scalar states show very little mixing, and correspondingly
the string tension is very close to its pure gauge value. 
The pure gauge quantities are furthermore insensitive to variations
of the scalar parameters in the action. All these findings are completely
analogous to the situation in the SU(2) fundamental Higgs model \cite{ptw}.
We conclude that the approximate decoupling of the pure gauge sector from 
the scalar sector is a very robust phenomenon in three dimensions, for
which an explanation is lacking at present.

Regarding its r{\^o}le as effective theory, we find that dimensional
reduction works remarkably well down to temperatures of $2T_c$, where
the lowest screening masses computed in the effective and the full
theories agree quantitatively.  This agreement is rather striking for
a temperature so close to the transition.  These findings confirm the
SU(2) adjoint Higgs model as the correct effective high temperature
theory to describe the thermodynamics of four-dimensional SU(2) pure
gauge theory.  In particular, our investigation has established
unambiguously that the $A_0$ degrees of freedom for temperatures not
too far above the phase transition constitute the lightest state in
the three-dimensional effective theory, and hence may not be
integrated out, as already discussed in \cite{ad,lp2}.  It furthermore
implies that the na\"{\i}ve hierarchy of scales $gT \ll g^2T $ does
not translate into a corresponding separation of the dynamics of the
$A_0$ and $A_i$ in this temperature range.

We expect no qualitative changes when moving from SU(2) to SU(3).  An
analogous investigation for SU(3) and three flavours of quarks poses
no additional complications and appears desirable in the light of our
results.

\subsection*{Acknowledgments}

We thank UKQCD for computer time on one of their workstations in
Edinburgh where part of these computations were performed, and
M.~Laine for a critical reading of the manuscript.
This work was supported in part by United Kingdom PPARC grant
GR/L22744.

\newpage

\end{document}

%% file: m_over_t.tex
\setlength{\unitlength}{0.240900pt}
\ifx\plotpoint\undefined\newsavebox{\plotpoint}\fi
\sbox{\plotpoint}{\rule[-0.200pt]{0.400pt}{0.400pt}}%
\begin{picture}(1875,900)(0,0)
\font\gnuplot=cmr10 at 10pt
\gnuplot
\sbox{\plotpoint}{\rule[-0.200pt]{0.400pt}{0.400pt}}%
\put(220.0,68.0){\rule[-0.200pt]{383.272pt}{0.400pt}}
\put(198,68){\makebox(0,0)[r]{{\large $0$}}}
\put(200.0,68.0){\rule[-0.200pt]{4.818pt}{0.400pt}}
\put(198,338){\makebox(0,0)[r]{{\large $5$}}}
\put(200.0,338.0){\rule[-0.200pt]{4.818pt}{0.400pt}}
\put(198,607){\makebox(0,0)[r]{{\large $10$}}}
\put(200.0,607.0){\rule[-0.200pt]{4.818pt}{0.400pt}}
\put(198,877){\makebox(0,0)[r]{{\large $15$}}}
\put(200.0,877.0){\rule[-0.200pt]{4.818pt}{0.400pt}}
\put(256,23){\makebox(0,0){$\scriptstyle 0^+_+$}}
\put(256.0,48.0){\rule[-0.200pt]{0.400pt}{4.818pt}}
\put(341,23){\makebox(0,0){$\scriptstyle 0^-_-$}}
\put(341.0,48.0){\rule[-0.200pt]{0.400pt}{4.818pt}}
\put(298,23){\makebox(0,0){$\scriptstyle 0^-_+$}}
\put(298.0,48.0){\rule[-0.200pt]{0.400pt}{4.818pt}}
\put(412,23){\makebox(0,0){$\scriptstyle 1^+_+$}}
\put(412.0,48.0){\rule[-0.200pt]{0.400pt}{4.818pt}}
\put(454,23){\makebox(0,0){$\scriptstyle 1^-_+$}}
\put(454.0,48.0){\rule[-0.200pt]{0.400pt}{4.818pt}}
\put(497,23){\makebox(0,0){$\scriptstyle 1^+_-$}}
\put(497.0,48.0){\rule[-0.200pt]{0.400pt}{4.818pt}}
\put(540,23){\makebox(0,0){$\scriptstyle 1^-_-$}}
\put(540.0,48.0){\rule[-0.200pt]{0.400pt}{4.818pt}}
\put(611,23){\makebox(0,0){$\scriptstyle 2^+_+$}}
\put(611.0,48.0){\rule[-0.200pt]{0.400pt}{4.818pt}}
\put(653,23){\makebox(0,0){$\scriptstyle 2^-_+$}}
\put(653.0,48.0){\rule[-0.200pt]{0.400pt}{4.818pt}}
\put(696,23){\makebox(0,0){$\scriptstyle 2^+_-$}}
\put(696.0,48.0){\rule[-0.200pt]{0.400pt}{4.818pt}}
\put(795,23){\makebox(0,0){$\scriptstyle 0^+_+$}}
\put(795.0,48.0){\rule[-0.200pt]{0.400pt}{4.818pt}}
\put(881,23){\makebox(0,0){$\scriptstyle 0^-_-$}}
\put(881.0,48.0){\rule[-0.200pt]{0.400pt}{4.818pt}}
\put(838,23){\makebox(0,0){$\scriptstyle 0^-_+$}}
\put(838.0,48.0){\rule[-0.200pt]{0.400pt}{4.818pt}}
\put(952,23){\makebox(0,0){$\scriptstyle 1^+_+$}}
\put(952.0,48.0){\rule[-0.200pt]{0.400pt}{4.818pt}}
\put(994,23){\makebox(0,0){$\scriptstyle 1^-_+$}}
\put(994.0,48.0){\rule[-0.200pt]{0.400pt}{4.818pt}}
\put(1037,23){\makebox(0,0){$\scriptstyle 1^+_-$}}
\put(1037.0,48.0){\rule[-0.200pt]{0.400pt}{4.818pt}}
\put(1079,23){\makebox(0,0){$\scriptstyle 1^-_-$}}
\put(1079.0,48.0){\rule[-0.200pt]{0.400pt}{4.818pt}}
\put(1150,23){\makebox(0,0){$\scriptstyle 2^+_+$}}
\put(1150.0,48.0){\rule[-0.200pt]{0.400pt}{4.818pt}}
\put(1193,23){\makebox(0,0){$\scriptstyle 2^-_+$}}
\put(1193.0,48.0){\rule[-0.200pt]{0.400pt}{4.818pt}}
\put(1236,23){\makebox(0,0){$\scriptstyle 2^+_-$}}
\put(1236.0,48.0){\rule[-0.200pt]{0.400pt}{4.818pt}}
\put(1335,23){\makebox(0,0){$\scriptstyle 0^+_+$}}
\put(1335.0,48.0){\rule[-0.200pt]{0.400pt}{4.818pt}}
\put(1420,23){\makebox(0,0){$\scriptstyle 0^-_-$}}
\put(1420.0,48.0){\rule[-0.200pt]{0.400pt}{4.818pt}}
\put(1378,23){\makebox(0,0){$\scriptstyle 0^-_+$}}
\put(1378.0,48.0){\rule[-0.200pt]{0.400pt}{4.818pt}}
\put(1491,23){\makebox(0,0){$\scriptstyle 1^+_+$}}
\put(1491.0,48.0){\rule[-0.200pt]{0.400pt}{4.818pt}}
\put(1534,23){\makebox(0,0){$\scriptstyle 1^-_+$}}
\put(1534.0,48.0){\rule[-0.200pt]{0.400pt}{4.818pt}}
\put(1577,23){\makebox(0,0){$\scriptstyle 1^+_-$}}
\put(1577.0,48.0){\rule[-0.200pt]{0.400pt}{4.818pt}}
\put(1619,23){\makebox(0,0){$\scriptstyle 1^-_-$}}
\put(1619.0,48.0){\rule[-0.200pt]{0.400pt}{4.818pt}}
\put(1690,23){\makebox(0,0){$\scriptstyle 2^+_+$}}
\put(1690.0,48.0){\rule[-0.200pt]{0.400pt}{4.818pt}}
\put(1733,23){\makebox(0,0){$\scriptstyle 2^-_+$}}
\put(1733.0,48.0){\rule[-0.200pt]{0.400pt}{4.818pt}}
\put(1775,23){\makebox(0,0){$\scriptstyle 2^+_-$}}
\put(1775.0,48.0){\rule[-0.200pt]{0.400pt}{4.818pt}}
\put(220.0,68.0){\rule[-0.200pt]{383.272pt}{0.400pt}}
\put(1811.0,68.0){\rule[-0.200pt]{0.400pt}{194.888pt}}
\put(220.0,877.0){\rule[-0.200pt]{383.272pt}{0.400pt}}
\put(155,472){\makebox(0,0){{\Large $\frac{M}{T}$}}}
\put(362,823){\makebox(0,0)[l]{{\large $T = 2T_c$}}}
\put(859,823){\makebox(0,0)[l]{{\large $T = 4T_c$}}}
\put(1356,823){\makebox(0,0)[l]{{\large $T = 5T_c$}}}
\put(220.0,68.0){\rule[-0.200pt]{0.400pt}{194.888pt}}
\put(256,403){\circle*{24}}
\put(256,551){\circle*{24}}
\put(795,304){\circle*{24}}
\put(795,454){\circle*{24}}
\put(1335,313){\circle*{24}}
\put(1335,414){\circle*{24}}
\put(256.0,399.0){\rule[-0.200pt]{0.400pt}{2.168pt}}
\put(246.0,399.0){\rule[-0.200pt]{4.818pt}{0.400pt}}
\put(246.0,408.0){\rule[-0.200pt]{4.818pt}{0.400pt}}
\put(256.0,543.0){\rule[-0.200pt]{0.400pt}{4.095pt}}
\put(246.0,543.0){\rule[-0.200pt]{4.818pt}{0.400pt}}
\put(246.0,560.0){\rule[-0.200pt]{4.818pt}{0.400pt}}
\put(795.0,301.0){\rule[-0.200pt]{0.400pt}{1.204pt}}
\put(785.0,301.0){\rule[-0.200pt]{4.818pt}{0.400pt}}
\put(785.0,306.0){\rule[-0.200pt]{4.818pt}{0.400pt}}
\put(795.0,448.0){\rule[-0.200pt]{0.400pt}{2.891pt}}
\put(785.0,448.0){\rule[-0.200pt]{4.818pt}{0.400pt}}
\put(785.0,460.0){\rule[-0.200pt]{4.818pt}{0.400pt}}
\put(1335.0,299.0){\rule[-0.200pt]{0.400pt}{6.745pt}}
\put(1325.0,299.0){\rule[-0.200pt]{4.818pt}{0.400pt}}
\put(1325.0,327.0){\rule[-0.200pt]{4.818pt}{0.400pt}}
\put(1335.0,375.0){\rule[-0.200pt]{0.400pt}{19.031pt}}
\put(1325.0,375.0){\rule[-0.200pt]{4.818pt}{0.400pt}}
\put(1325.0,454.0){\rule[-0.200pt]{4.818pt}{0.400pt}}
\put(256,222){\raisebox{-.8pt}{\makebox(0,0){$\Box$}}}
\put(256,481){\raisebox{-.8pt}{\makebox(0,0){$\Box$}}}
\put(341,313){\raisebox{-.8pt}{\makebox(0,0){$\Box$}}}
\put(341,512){\raisebox{-.8pt}{\makebox(0,0){$\Box$}}}
\put(298,597){\raisebox{-.8pt}{\makebox(0,0){$\Box$}}}
\put(795,224){\raisebox{-.8pt}{\makebox(0,0){$\Box$}}}
\put(795,424){\raisebox{-.8pt}{\makebox(0,0){$\Box$}}}
\put(881,287){\raisebox{-.8pt}{\makebox(0,0){$\Box$}}}
\put(881,448){\raisebox{-.8pt}{\makebox(0,0){$\Box$}}}
\put(838,644){\raisebox{-.8pt}{\makebox(0,0){$\Box$}}}
\put(1335,224){\raisebox{-.8pt}{\makebox(0,0){$\Box$}}}
\put(1335,414){\raisebox{-.8pt}{\makebox(0,0){$\Box$}}}
\put(1420,275){\raisebox{-.8pt}{\makebox(0,0){$\Box$}}}
\put(1420,424){\raisebox{-.8pt}{\makebox(0,0){$\Box$}}}
\put(1378,644){\raisebox{-.8pt}{\makebox(0,0){$\Box$}}}
\put(256.0,220.0){\rule[-0.200pt]{0.400pt}{0.964pt}}
\put(246.0,220.0){\rule[-0.200pt]{4.818pt}{0.400pt}}
\put(246.0,224.0){\rule[-0.200pt]{4.818pt}{0.400pt}}
\put(256.0,475.0){\rule[-0.200pt]{0.400pt}{2.891pt}}
\put(246.0,475.0){\rule[-0.200pt]{4.818pt}{0.400pt}}
\put(246.0,487.0){\rule[-0.200pt]{4.818pt}{0.400pt}}
\put(341.0,311.0){\rule[-0.200pt]{0.400pt}{1.204pt}}
\put(331.0,311.0){\rule[-0.200pt]{4.818pt}{0.400pt}}
\put(331.0,316.0){\rule[-0.200pt]{4.818pt}{0.400pt}}
\put(341.0,505.0){\rule[-0.200pt]{0.400pt}{3.373pt}}
\put(331.0,505.0){\rule[-0.200pt]{4.818pt}{0.400pt}}
\put(331.0,519.0){\rule[-0.200pt]{4.818pt}{0.400pt}}
\put(298.0,573.0){\rule[-0.200pt]{0.400pt}{11.322pt}}
\put(288.0,573.0){\rule[-0.200pt]{4.818pt}{0.400pt}}
\put(288.0,620.0){\rule[-0.200pt]{4.818pt}{0.400pt}}
\put(795.0,223.0){\rule[-0.200pt]{0.400pt}{0.723pt}}
\put(785.0,223.0){\rule[-0.200pt]{4.818pt}{0.400pt}}
\put(785.0,226.0){\rule[-0.200pt]{4.818pt}{0.400pt}}
\put(795.0,419.0){\rule[-0.200pt]{0.400pt}{2.409pt}}
\put(785.0,419.0){\rule[-0.200pt]{4.818pt}{0.400pt}}
\put(785.0,429.0){\rule[-0.200pt]{4.818pt}{0.400pt}}
\put(881.0,285.0){\rule[-0.200pt]{0.400pt}{0.964pt}}
\put(871.0,285.0){\rule[-0.200pt]{4.818pt}{0.400pt}}
\put(871.0,289.0){\rule[-0.200pt]{4.818pt}{0.400pt}}
\put(881.0,443.0){\rule[-0.200pt]{0.400pt}{2.650pt}}
\put(871.0,443.0){\rule[-0.200pt]{4.818pt}{0.400pt}}
\put(871.0,454.0){\rule[-0.200pt]{4.818pt}{0.400pt}}
\put(838.0,624.0){\rule[-0.200pt]{0.400pt}{9.636pt}}
\put(828.0,624.0){\rule[-0.200pt]{4.818pt}{0.400pt}}
\put(828.0,664.0){\rule[-0.200pt]{4.818pt}{0.400pt}}
\put(1335.0,219.0){\rule[-0.200pt]{0.400pt}{2.650pt}}
\put(1325.0,219.0){\rule[-0.200pt]{4.818pt}{0.400pt}}
\put(1325.0,230.0){\rule[-0.200pt]{4.818pt}{0.400pt}}
\put(1335.0,386.0){\rule[-0.200pt]{0.400pt}{13.490pt}}
\put(1325.0,386.0){\rule[-0.200pt]{4.818pt}{0.400pt}}
\put(1325.0,442.0){\rule[-0.200pt]{4.818pt}{0.400pt}}
\put(1420.0,265.0){\rule[-0.200pt]{0.400pt}{4.818pt}}
\put(1410.0,265.0){\rule[-0.200pt]{4.818pt}{0.400pt}}
\put(1410.0,285.0){\rule[-0.200pt]{4.818pt}{0.400pt}}
\put(1420.0,406.0){\rule[-0.200pt]{0.400pt}{8.672pt}}
\put(1410.0,406.0){\rule[-0.200pt]{4.818pt}{0.400pt}}
\put(1410.0,442.0){\rule[-0.200pt]{4.818pt}{0.400pt}}
\put(1378.0,624.0){\rule[-0.200pt]{0.400pt}{9.636pt}}
\put(1368.0,624.0){\rule[-0.200pt]{4.818pt}{0.400pt}}
\put(1368.0,664.0){\rule[-0.200pt]{4.818pt}{0.400pt}}
\put(412,696){\makebox(0,0){$\triangle$}}
\put(454,672){\makebox(0,0){$\triangle$}}
\put(497,479){\makebox(0,0){$\triangle$}}
\put(540,464){\makebox(0,0){$\triangle$}}
\put(952,643){\makebox(0,0){$\triangle$}}
\put(994,573){\makebox(0,0){$\triangle$}}
\put(1037,399){\makebox(0,0){$\triangle$}}
\put(1079,408){\makebox(0,0){$\triangle$}}
\put(1491,626){\makebox(0,0){$\triangle$}}
\put(1534,565){\makebox(0,0){$\triangle$}}
\put(1577,399){\makebox(0,0){$\triangle$}}
\put(1619,387){\makebox(0,0){$\triangle$}}
\put(412.0,671.0){\rule[-0.200pt]{0.400pt}{12.045pt}}
\put(402.0,671.0){\rule[-0.200pt]{4.818pt}{0.400pt}}
\put(402.0,721.0){\rule[-0.200pt]{4.818pt}{0.400pt}}
\put(454.0,651.0){\rule[-0.200pt]{0.400pt}{10.118pt}}
\put(444.0,651.0){\rule[-0.200pt]{4.818pt}{0.400pt}}
\put(444.0,693.0){\rule[-0.200pt]{4.818pt}{0.400pt}}
\put(497.0,473.0){\rule[-0.200pt]{0.400pt}{2.891pt}}
\put(487.0,473.0){\rule[-0.200pt]{4.818pt}{0.400pt}}
\put(487.0,485.0){\rule[-0.200pt]{4.818pt}{0.400pt}}
\put(540.0,454.0){\rule[-0.200pt]{0.400pt}{4.818pt}}
\put(530.0,454.0){\rule[-0.200pt]{4.818pt}{0.400pt}}
\put(530.0,474.0){\rule[-0.200pt]{4.818pt}{0.400pt}}
\put(952.0,603.0){\rule[-0.200pt]{0.400pt}{19.272pt}}
\put(942.0,603.0){\rule[-0.200pt]{4.818pt}{0.400pt}}
\put(942.0,683.0){\rule[-0.200pt]{4.818pt}{0.400pt}}
\put(994.0,555.0){\rule[-0.200pt]{0.400pt}{8.672pt}}
\put(984.0,555.0){\rule[-0.200pt]{4.818pt}{0.400pt}}
\put(984.0,591.0){\rule[-0.200pt]{4.818pt}{0.400pt}}
\put(1037.0,394.0){\rule[-0.200pt]{0.400pt}{2.168pt}}
\put(1027.0,394.0){\rule[-0.200pt]{4.818pt}{0.400pt}}
\put(1027.0,403.0){\rule[-0.200pt]{4.818pt}{0.400pt}}
\put(1079.0,403.0){\rule[-0.200pt]{0.400pt}{2.168pt}}
\put(1069.0,403.0){\rule[-0.200pt]{4.818pt}{0.400pt}}
\put(1069.0,412.0){\rule[-0.200pt]{4.818pt}{0.400pt}}
\put(1491.0,593.0){\rule[-0.200pt]{0.400pt}{15.899pt}}
\put(1481.0,593.0){\rule[-0.200pt]{4.818pt}{0.400pt}}
\put(1481.0,659.0){\rule[-0.200pt]{4.818pt}{0.400pt}}
\put(1534.0,547.0){\rule[-0.200pt]{0.400pt}{8.431pt}}
\put(1524.0,547.0){\rule[-0.200pt]{4.818pt}{0.400pt}}
\put(1524.0,582.0){\rule[-0.200pt]{4.818pt}{0.400pt}}
\put(1577.0,392.0){\rule[-0.200pt]{0.400pt}{3.373pt}}
\put(1567.0,392.0){\rule[-0.200pt]{4.818pt}{0.400pt}}
\put(1567.0,406.0){\rule[-0.200pt]{4.818pt}{0.400pt}}
\put(1619.0,381.0){\rule[-0.200pt]{0.400pt}{2.891pt}}
\put(1609.0,381.0){\rule[-0.200pt]{4.818pt}{0.400pt}}
\put(1609.0,393.0){\rule[-0.200pt]{4.818pt}{0.400pt}}
\put(611,615){\circle*{24}}
\put(1150,504){\circle*{24}}
\put(1690,475){\circle*{24}}
\put(611.0,605.0){\rule[-0.200pt]{0.400pt}{5.059pt}}
\put(601.0,605.0){\rule[-0.200pt]{4.818pt}{0.400pt}}
\put(601.0,626.0){\rule[-0.200pt]{4.818pt}{0.400pt}}
\put(1150.0,496.0){\rule[-0.200pt]{0.400pt}{4.095pt}}
\put(1140.0,496.0){\rule[-0.200pt]{4.818pt}{0.400pt}}
\put(1140.0,513.0){\rule[-0.200pt]{4.818pt}{0.400pt}}
\put(1690.0,464.0){\rule[-0.200pt]{0.400pt}{5.059pt}}
\put(1680.0,464.0){\rule[-0.200pt]{4.818pt}{0.400pt}}
\put(1680.0,485.0){\rule[-0.200pt]{4.818pt}{0.400pt}}
\put(611,543){\circle{24}}
\put(653,546){\circle{24}}
\put(696,588){\circle{24}}
\put(1150,469){\circle{24}}
\put(1193,474){\circle{24}}
\put(1236,494){\circle{24}}
\put(1690,451){\circle{24}}
\put(1733,469){\circle{24}}
\put(1775,478){\circle{24}}
\put(611.0,535.0){\rule[-0.200pt]{0.400pt}{4.095pt}}
\put(601.0,535.0){\rule[-0.200pt]{4.818pt}{0.400pt}}
\put(601.0,552.0){\rule[-0.200pt]{4.818pt}{0.400pt}}
\put(653.0,536.0){\rule[-0.200pt]{0.400pt}{4.818pt}}
\put(643.0,536.0){\rule[-0.200pt]{4.818pt}{0.400pt}}
\put(643.0,556.0){\rule[-0.200pt]{4.818pt}{0.400pt}}
\put(696.0,577.0){\rule[-0.200pt]{0.400pt}{5.541pt}}
\put(686.0,577.0){\rule[-0.200pt]{4.818pt}{0.400pt}}
\put(686.0,600.0){\rule[-0.200pt]{4.818pt}{0.400pt}}
\put(1150.0,463.0){\rule[-0.200pt]{0.400pt}{2.891pt}}
\put(1140.0,463.0){\rule[-0.200pt]{4.818pt}{0.400pt}}
\put(1140.0,475.0){\rule[-0.200pt]{4.818pt}{0.400pt}}
\put(1193.0,468.0){\rule[-0.200pt]{0.400pt}{3.132pt}}
\put(1183.0,468.0){\rule[-0.200pt]{4.818pt}{0.400pt}}
\put(1183.0,481.0){\rule[-0.200pt]{4.818pt}{0.400pt}}
\put(1236.0,487.0){\rule[-0.200pt]{0.400pt}{3.613pt}}
\put(1226.0,487.0){\rule[-0.200pt]{4.818pt}{0.400pt}}
\put(1226.0,502.0){\rule[-0.200pt]{4.818pt}{0.400pt}}
\put(1690.0,419.0){\rule[-0.200pt]{0.400pt}{15.658pt}}
\put(1680.0,419.0){\rule[-0.200pt]{4.818pt}{0.400pt}}
\put(1680.0,484.0){\rule[-0.200pt]{4.818pt}{0.400pt}}
\put(1733.0,460.0){\rule[-0.200pt]{0.400pt}{4.336pt}}
\put(1723.0,460.0){\rule[-0.200pt]{4.818pt}{0.400pt}}
\put(1723.0,478.0){\rule[-0.200pt]{4.818pt}{0.400pt}}
\put(1775.0,464.0){\rule[-0.200pt]{0.400pt}{6.745pt}}
\put(1765.0,464.0){\rule[-0.200pt]{4.818pt}{0.400pt}}
\put(1765.0,492.0){\rule[-0.200pt]{4.818pt}{0.400pt}}
\end{picture}

%% file: su2ha_spec_2.bbl
\begin{thebibliography}{99}

\bibitem{dr}
P. Ginsparg,
Nucl. Phys. B 170 (1980) 388;\\
T. Appelquist and R. Pisarski,
Phys. Rev. D 23 (1981) 2305.

\bibitem{ewn}
K. Kajantie, M. Laine, K. Rummukainen and M. Shaposhnikov,
Phys. Rev. Lett. 77 (1996) 2887 [hep-ph/9605028];\\
%
M. G\"urtler, E.-M. Ilgenfritz and A. Schiller, 
Phys. Rev. D 56 (1997) 3888 [hep-lat/9704013];\\
%
F. Karsch, T. Neuhaus, A. Patk\'os and J. Rank, 
Nucl. Phys. Proc. Suppl. 53 (1997) 623 [hep-lat/9608087].

\bibitem{ewa}
W. Buchm\"uller and O. Philipsen, Nucl. Phys. B 443 (1995) 47;
Phys. Lett. B 397 (1997) 112 [hep-ph/9612286];\\
%
N. Tetradis, Nucl. Phys. B 488 (1997) 92 [hep-ph/9608272].

\bibitem{4dew}
F. Csikor, Z. Fodor, J. Heitger, 
Phys. Rev. Lett. 82 (1999) 21 [hep-ph/9809291];\\
%
M. Laine, JHEP 9906 (1999) 020 [hep-ph/9903513].

\bibitem{ad}
K. Kajantie, M. Laine, K. Rummukainen and M. Shaposhnikov,
Nucl. Phys. B 503 (1997) 357 [hep-ph/9704416].

\bibitem{bra}
E. Braaten and A. Nieto, Phys. Rev. Lett. 76 (1996) 1417; 
Phys. Rev. D 53 (1996) 3421 [hep-ph/9510408].

\bibitem{usad}
A. Hart, O. Philipsen, M. Teper and J. Stack, 
Phys. Lett. B 396 (1997) 217 [hep-lat/9612021].

\bibitem{4dpt}
J. Fingberg, U. Heller and F. Karsch, 
Nucl. Phys. B 392 (1993) 493 [hep-lat/9208012].

\bibitem{dg}
S. Datta and S. Gupta,
Nucl. Phys. B 534 (1998) 392 [hep-lat/9806034]. 

\bibitem{dg2}
S. Datta and S. Gupta, [hep-lat/9906023].

\bibitem{rei1}
P. Lacock, D. Miller and T. Reisz, Nucl. Phys. B 369 (1992) 501.

\bibitem{kar1}
F. Karsch, M. Oevers and P. Petreczky, Phys. Lett. B 442 (1998) 291;\\
U. Heller, F. Karsch and J. Rank, Phys. Rev. D 57 (1998) 1438.

\bibitem{rei}
P. Lacock, D.E. Miller, B. Petersson
and T. Reisz, Nucl. Phys. B 418 (1994) 3 [hep-lat/9310014].
 
\bibitem{mt}
M. Teper,
Phys. Rev. D 59 (1999) 014512 [hep-lat/9804008].

\bibitem{ptw}
O. Philipsen, M. Teper and H. Wittig,
Nucl. Phys. B 469 (1996) 445 [hep-lat/9602006];\\
Nucl. Phys. B 528 (1998) 379 [hep-lat/9709145]

\bibitem{lai95}
M. Laine, Nucl. Phys. B 451 (1995) 484 [hep-lat/9504001].

\bibitem{block}
M. Teper, Phys. Lett. B 187 (1987) 345.

\bibitem{cm87a}
C. Michael, J. Phys. G 13 (1987) 1001.

\bibitem{ud}
K. Fabricius and O. Haan, Phys. Lett. B 143 (1984) 459;\\
A.D. Kennedy and B.J. Pendleton, Phys. Lett. B 156 (1985) 393.

\bibitem{bunk}
B. Bunk, Nucl. Phys. B (Proc. Suppl.) 42 (1995) 566.

\bibitem{moore}
G. D. Moore, Nucl. Phys. B 523 (1998) 569 [hep-lat/9709053].

\bibitem{sb}
O. Philipsen and H. Wittig,
Phys. Rev. Lett. 81 (1998) 4056 [hep-lat/9807020].

\bibitem{for85}
P. de Forcrand, G. Schierholz, H. Schneider and M. Teper,
Phys. Lett. 160 B (1985) 137.

\bibitem{ay}
P. Arnold and L. Yaffe,
Phys. Rev. D 52 (1995) 7208 [hep-ph/9508280].

\bibitem{lo}
E. Shuryak, Zh. Eksp. Teor. Fiz. 74 (1978) 408 
[Sov. Phys. JETP 47 (1978) 212];\\
J. Kapusta, Nucl. Phys. B 148 (1979) 461;\\
D. Gross, R. Pisarski and L. Yaffe, Rev. Mod. Phys. 53 (1981) 43.

\bibitem{rebhan} 
A.K. Rebhan,
Phys. Rev. D 48 (1993) R3967 [hep-ph/9308232];\\
Nucl. Phys. B 430 (1994) 319 [hep-ph/9408262].

\bibitem{lp1}
M. Laine and O. Philipsen,
Nucl. Phys. B 523 (1998) 267 [hep-lat/9711022].

\bibitem{lp2}
M. Laine and O. Philipsen, Phys. Lett. B 459 (1999) 259 [hep-lat/9905004].

\bibitem{mD}
K. Kajantie, M. Laine, J. Peisa, A. Rajantie, K. Rummukainen and
M. Shaposhnikov,
Phys. Rev. Lett. 79 (1997) 3130 [hep-ph/9708207].

\end{thebibliography}
